\documentclass[10pt,twocolumn,twoside]{IEEEtran}

\usepackage{setspace}
\usepackage{amsmath}
\usepackage{algorithm}
\usepackage{graphicx}
\usepackage{color}
\usepackage{epsfig}
\usepackage{subfigure}
\usepackage{amsfonts}
\usepackage{amssymb}

\title{Sensing Matrix Optimization for Block-Sparse Decoding
\thanks{The authors are with the Technion - Israel Institute of
Technology, Haifa, Israel. Email: kevin@tx.technion.ac.il,
lihi@ee.technion.ac.il, yonina@ee.technion.ac.il.}}

\author{Kevin Rosenblum, Lihi Zelnik-Manor, Yonina C. Eldar}

\begin{document}

\maketitle

\begin{abstract}
Recent work has demonstrated that using a carefully designed sensing
matrix rather than a random one, can improve the performance of
compressed sensing. In particular, a well-designed sensing matrix
can reduce the coherence between the atoms of the equivalent
dictionary, and as a consequence, reduce the reconstruction error.
In some applications, the signals of interest can be well
approximated by a union of a small number of subspaces (e.g., face
recognition and motion segmentation). This implies the existence of
a dictionary which leads to {\em block-sparse} representations. In
this work, we propose a framework for sensing matrix design that
improves the ability of block-sparse approximation techniques to
reconstruct and classify signals. This method is based on minimizing
a weighted sum of the inter-block coherence and the sub-block
coherence of the equivalent dictionary. Our experiments show that
the proposed algorithm significantly improves signal recovery and
classification ability of the \emph{Block-OMP} algorithm compared to
sensing matrix optimization methods that do not employ block
structure.
\end{abstract}

%
%

\section{Introduction}
\label{sec:intro}

The framework of compressed sensing aims at recovering an unknown
vector $x \in R^N$ from an under-determined system of linear
equations $y = Ax$, where $A\in R^{M \times N}$ is a sensing matrix,
and $y \in R^M$ is an observation vector with $M<N$. Since the
system is under-determined, $x$ can not be recovered without
additional information. In \cite{CS1,CS2} it was shown that when $x$
is known to have a sufficiently sparse representation, and when $A$
is randomly generated, $x$ can be recovered uniquely with high
probability from the measurements $y$. More specifically, the
assumption is that $x$ can be represented as $x=D\theta$ for some
orthogonal dictionary $D\in R^{N \times N}$, where $\theta \in R^N$
is sufficiently sparse. The vector $x$ can then be recovered
regardless of $D$ and irrespective of the locations of the nonzero
entries of $\theta$. This can be achieved by approximating the
sparsest representation $\theta$ using methods such as \emph{Basis
Pursuit} (BP) \cite{BP, CS1} and \emph{Orthogonal Matching Pursuit}
(OMP) \cite{OMP1,OMP2}. In practice, overcomplete dictionaries $D\in
R^{N \times K}$ with $K \geq N$ lead to improved sparse
representations and are better suited for most applications.
Therefore, we treat the more general case of overcomplete
dictionaries in this paper.

A simple way to characterize the recovery ability of sparse
approximation algorithms was presented in \cite{OMP1}, using the
coherence between the columns of the equivalent dictionary $E=AD$.
When the coherence is sufficiently low, OMP and BP are guaranteed to
recover the sparse vector $\theta$. Accordingly, recent work
\cite{Sens1,Sens2,sapiro} has demonstrated that designing a sensing
matrix such that the coherence of $E$ is low improves the ability to
recover $\theta$. The proposed methods yield good results for
general sparse vectors.

In some applications, however, the representations have a unique
sparsity structure that can be exploited. Our interest is in the
case of signals that are drawn from a union of a small number of
subspaces \cite{do,davies,kfir,BBP1}. This occurs naturally, for
example, in face recognition \cite{face1,face2}, motion segmentation
\cite{motion}, multi-band signals \cite{mult1,mult2,mult3},
measurements of gene expression levels \cite{DNA}, and more. For
such signals, sorting the dictionary atoms according to the
underlying subspaces leads to sparse representations which exhibit a
block-sparse structure, i.e., the nonzero coefficients in $\theta$
occur in clusters of varying sizes. Several methods, such as
\emph{Block-BP} (BBP) \cite{BBP1,BBP2,BBP3} and \emph{Block-OMP}
(BOMP) \cite{BOMP1,BOMP2} have been proposed to take advantage of
this block structure in recovering the block-sparse representations
$\theta$. Bounds on the recovery performance were presented in
\cite{BBP1} based on the block restricted isometry property (RIP),
and in \cite{BOMP1} using appropriate coherence measures. In
particular, it was shown in \cite{BOMP1} that under conditions on
the \emph{inter-block coherence} (i.e., the maximal coherence
between two blocks) and the \emph{sub-block coherence} (i.e., the
maximal coherence between two atoms in the same block) of the
equivalent dictionary $E$, Block-OMP is guaranteed to recover the
block-sparse vector $\theta$.

In this paper we propose a method for designing a sensing matrix,
assuming that a block-sparsifying dictionary is provided. A method
for learning a block-sparsifying dictionary is developed in
\cite{BKSVDSAC}. Our approach improves the recovery ability of
block-sparse approximation algorithms by targeting the Gram matrix
of the equivalent dictionary, an approach similar in spirit to that
of \cite{Sens2,sapiro}. While \cite{Sens2} and \cite{sapiro}
targeted minimization of the coherence between atoms, our method,
which will be referred to as \emph{Weighted Coherence Minimization}
(WCM), aims at reducing a weighted sum of the inter-block coherence
and the sub-block coherence.

It turns out that the weighted coherence objective is hard to
minimize directly. To derive an efficient algorithm, we use the
bound-optimization method, and replace our objective with an easier
to minimize surrogate function that is updated in each optimization
step \cite{BO}. We develop a closed form solution for minimizing the
surrogate function in each step, and prove that its iterative
minimization is guaranteed to converge to a local solution of the
original problem.

Our experiments reveal that minimizing the sub-block coherence is
more important than minimizing the inter-block coherence. By giving
more weight to minimizing the sub-block coherence, the proposed
algorithm yields sensing matrices that lead to equivalent
dictionaries with nearly orthonormal blocks. Simulations show that
such sensing matrices significantly improve signal reconstruction
and signal classification results compared to previous approaches
that do not employ block structure.

We begin by reviewing previous work on sensing matrix design in
Section~\ref{sec:prior-work}. In Section~\ref{sec:sensing} we
introduce our definitions of total inter-block coherence and total
sub-block coherence. We then present the objective for sensing
matrix design, and show that it can be considered as a direct
extension of the one used in \cite{sapiro} to the case of blocks. We
present the WCM algorithm for minimizing the proposed objective in
Section~\ref{sec:WCM} and prove its convergence in Appendix A. We
evaluate the performance of the proposed algorithm and compare it to
previous work in Section~\ref{sec:experiments}.

Throughout the paper, we denote vectors by lowercase letters, e.g.,
$x$, and matrices by uppercase letters, e.g., $A$. $A'$ is the
transpose of $A$. The $j$th column of the matrix $A$ is $A_j$, and
the $i$th row is $A^i$. The entry of $A$ in the row with index $i$
and the column with index $j$ is $A^i_j$. We define the Frobenius
norm by $\|A\|_F\equiv\sqrt{\sum_j{\|A_j\|_2^2}}$, and the
$l_p$-norm of a vector $x$ by $\|x\|_p$. The $l_0$-norm $\|x\|_0$
counts the number of non-zero entries in $x$. We denote the identity
matrix by $I$ or $I_s$ when the dimension is not clear from the
context. The largest eigenvalue of the positive-semidefinite matrix
$B'B$ is written as $\lambda_\textrm{max}(B)$.

%
%

\section{Prior work on sensing matrix design}
\label{sec:prior-work}

The goal of sensing matrix design is to construct a sensing matrix
$A\in R^{M \times N}$ with $M<N$ that improves the recovery ability
for a given sparsifying dictionary $D\in R^{N \times K}$ with $K\geq
N$. In other words, $A$ is designed to improve the ability of sparse
approximation algorithms such as BP and OMP to recover the sparsest
representation $\theta$ from
\begin{equation}
y=AD\theta=E\theta,
\end{equation}
where $E$ is the equivalent dictionary. In this section we briefly
review the sensing matrix design method introduced by
Duarte-Carvajalino and Sapiro \cite{sapiro}. Their algorithm was
shown to provide significant improvement in reconstruction success.

The motivation to design sensing matrices stems from the theoretical
work of \cite{OMP1}, where it was shown that BP and OMP succeed in
recovering $\theta$ when the following condition holds:
\begin{equation}
\|\theta\|_0 \leq \frac{1}{2}\left(1+ \frac{1}{\mu}\right).
\label{bound1}
\end{equation}
Here $\mu$ is the coherence defined by:
\begin{equation}
\mu \equiv \max_{i \neq j}\frac{|E_i'E_j|}{\|E_i\|_2\|E_j\|_2}.
\end{equation}
The smaller $\mu$, the higher the bound on the sparsity of $\theta$.
Since $E$ is overcomplete, and as a consequence not orthogonal,
$\mu$ will always be strictly positive. Condition \eqref{bound1} is
a worst-case bound and does not reflect the average recovery ability
of sparse approximation methods. However, it does suggest that
recovery may be improved when $E$ is as orthogonal as possible.

Motivated by these observations, Duarte-Carvajalino and Sapiro
\cite{sapiro} proposed designing a sensing matrix $A$ by minimizing
$\|E'E-I\|_F^2$. This problem can be written as:
\begin{equation}
    \min_{A}\|E'E-I\|_F^2=\min_{A}\|D'A'AD-I\|_F^2.
    \label{sap}
\end{equation}
It is important to note that rather than minimizing $\mu$,
\eqref{sap} minimizes the sum of the squared inner products of all
pairs of atoms in $E$, referred to as the total coherence $\mu^t$:
\begin{equation}
\mu^t=\sum_{j,i\neq j} (E_i'E_j)^2.
\end{equation}
At the same time, solving \eqref{sap} keeps the norms of the atoms
close to $1$.

While an approximate solution to \eqref{sap} has already been
presented in \cite{sapiro}, we provide an exact solution that will
be of use in the next sections. To solve \eqref{sap}, we rewrite its
objective using the well-known relation between the Frobenius norm
and the trace, $\|C\|_F^2=\textrm{tr}(CC')$:
\begin{align}
    \|E'E-I_K\|_F^2 =&\textrm{tr}(E'E E'E-2E'E+I_K)\nonumber\\
    =& \textrm{tr}(E E'E E'-2E E'+I_M)+(K-M)\nonumber\\
    =& \|E E'-I_M\|_F^2+(K-M)\nonumber\\
    =& \|ADD'A'-I_M\|_F^2+(K-M).
\label{sap2}
\end{align}
Since the first term in \eqref{sap2} is always positive, the
objective of \eqref{sap} is lower bounded by $\|E'E-I\|_F^2\geq
K-M$.

From \eqref{sap2} it follows that minimizing \eqref{sap} is
equivalent to the minimization of $\|ADD'A'-I_M\|_F^2$. A solution
to this problem can be achieved in closed form as follows. Let
$U\Lambda U'$ be the eigenvalue decomposition of $DD'$, and let
$\Gamma_{M\times N}= AU\Lambda^{1/2}$. Then, \eqref{sap} is
equivalent to:
\begin{equation}
    \min_{A}\|\Gamma\Gamma'-I\|_F^2.
    \label{sap3}
\end{equation}
This problem is solved by choosing $\Gamma$ to be any matrix with
orthonormal rows, such as $\Gamma=[I_M \textrm{ } 0]$, leading to
$\Gamma\Gamma'=I$. The optimal sensing matrix is then given by $A =
\Gamma \Lambda^{-1/2} U'$. Here, and throughout the paper, we assume
that $D$ has full row rank, guaranteeing that $\Lambda$ is
invertible. Note that the global minimum of the objective in
\eqref{sap} equals $K-M$. The benefits of using such a sensing
matrix were shown empirically in \cite{sapiro}.

The same solution is obtained by setting the derivative of
\eqref{sap3} equal to zero:
\begin{equation}
    \frac{\partial\|\Gamma\Gamma'-I\|_F^2}{\partial\Gamma}=4(\Gamma\Gamma'\Gamma-\Gamma)=0
    \label{sap4}
\end{equation}
It can be deduced from \eqref{sap4} that for stationary points, the
singular values of $\Gamma$ must be equal to either one or zero.
However, only when all the $M$ singular values of $\Gamma$ equal
one, i.e., $\Gamma$ has full row rank, we have a local minimum (the
other stationary points being a local maximum and saddle points). It
is important to keep in mind that even though the objective is not
convex, every local minimum is a global minimum as well.

\section{Sensing matrix design for block-sparse decoding}
\label{sec:sensing}

The design of a sensing matrix according to \cite{sapiro} does not
take advantage of block structure in the sparse representations of
the data. In this section we formulate the problem of sensing matrix
design for block-sparse decoding. We first introduce the basic
concepts of block-sparsity, and then present an objective which can
be seen as an extension of \eqref{sap} to the case of block-sparse
decoding.

\subsection{Block-sparse decoding}
\label{ssec:block}

The framework of block-sparse decoding aims at recovering an unknown
vector $x \in R^N$ from an under-determined system of linear
equations $y = Ax$, where $A\in R^{M \times N}$ is a sensing matrix,
and $y \in R^M$ is an observation vector with $M<N$. The difference
with sparse recovery lies in the assumption that $x$ has a
sufficiently block-sparse representation $\theta \in R^N$ with
respect to some orthogonal block-sparsifying dictionary $D\in R^{N
\times N}$. The vector $x$ can then be recovered by approximating
the block-sparsest representation corresponding to the measurements
$y$ using methods such as Block-BP (BBP) \cite{BBP1,BBP2,BBP3} and
Block-OMP (BOMP) \cite{BOMP1,BOMP2}.

A block-sparsifying dictionary $D$ is a dictionary whose atoms are
sorted in blocks which enable block-sparse representations for a set
of signals. We can represent $D$ as a concatenation of $B$
column-blocks $D[j]$ of size $N\times s_j$, where $s_j$ is the
number of atoms belonging to the $j$th block:
\begin{align*}
D = [D[1]\textrm{ }D[2]\textrm{ }\ldots\textrm{ }D[B]].
\end{align*}
Similarly, we view the representation $\theta$ as a concatenation of
$B$ blocks $\theta[j]$ of length $s_j$:
\begin{align*}
\theta = [\theta[1]\textrm{ }\theta[2]\textrm{ }\ldots\textrm{
}\theta[B]]'.
\end{align*}
We say that a representation $\theta$ is $k$-block-sparse if its
nonzero values are concentrated in $k$ blocks only. This is denoted
by $\|\theta\|_{2,0}\leq k$, where
\begin{align*}
\|\theta\|_{2,0} = \sum_{j=1}^B I(\|\theta[j]\|_2>0).
\end{align*}
The indicator function $I(\cdot)$ counts the number of blocks in
$\theta$ with nonzero Euclidean norm.

\subsection{Problem definition}
\label{ssec:problem}

For a given block-sparsifying dictionary $D\in R^{N \times K}$ with
$K\geq N$, we wish to design a sensing matrix $A\in R^{M \times N}$
that improves the recovery ability of block-sparse approximation
algorithms. Note that we allow $D$ to be overcomplete.

A performance bound on the recovery success of block-sparse signals
has been developed in \cite{BOMP1} for the case of a dictionary $D$
with blocks of a fixed size $s$ (i.e., $s_i = s_j = s$) and an
equivalent dictionary $E=AD$ with normalized columns. The bound is a
function of the Gram matrix $G \in R^{K \times K}$ of the equivalent
dictionary, defined as $E'E$. The $(i,j)$th block of $G$,
$E[i]'E[j]$, is denoted by $G[i,j] \in R^{s_i \times s_j}$. The
$(i,j)$th block of any other $K\times K$ matrix will be denoted
similarly. It was shown in \cite{BOMP1} that BBP and BOMP succeed in
recovering the block sparsest representation $\theta$ corresponding
to the measurements $y=E\theta$ when the following condition holds:
\begin{equation}
\label{bound2}
    \|\theta\|_{2,0}<\frac{1}{2s}\left(
\mu_B^{-1}+s-(s-1)\frac{\nu}{\mu_B}\right).
\end{equation}
Here \[\mu_B\equiv\max_{j, i \neq
j}\frac{1}{s}\sqrt{\lambda_\textrm{max}(G[i,j]'G[i,j])}\] is the
{\em inter-block coherence} and \[\nu \equiv \max_j\max_{n,m\neq
n}|(G[j,j])^m_n|\] is the {\em sub-block coherence}. The inter-block
coherence $\mu_B$ is a generalization of the coherence $\mu$, and
describes the global properties of the equivalent dictionary. More
specifically, $\mu_B$ measures the cosine of the minimal angle
between two blocks in $E$. The sub-block coherence $\nu$ describes
the local properties of the dictionary, by measuring the cosine of
the minimal angle between two atoms in the same block in $E$. Note,
that when $s=1$, \eqref{bound2} reduces to the bound in the sparse
case \eqref{bound1}. The term $\mu_B^{-1}$ in \eqref{bound2}
suggests that $\mu_B$ needs to be reduced in order to loosen the
bound. On the other hand, the term $-(s-1)\frac{\nu}{\mu_B}$ implies
that the ratio $\frac{\nu}{\mu_B}$ should be small. This leads to a
trade-off between minimizing $\mu_B$ and minimizing $\nu$ to loosen
the bound, which is reflected in the sensing matrix design objective
presented later in this section.

\begin{figure}
\centering
\includegraphics[width=100mm,angle=0,scale=0.5]{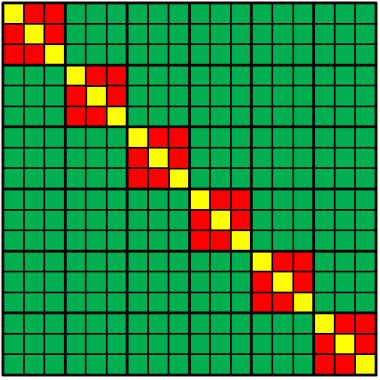}
\caption{A graphical depiction of the Gram matrix $G$ of an
equivalent dictionary $E$ with $6$ blocks of size $3$. The entries
belonging to different blocks are in green, the off-diagonal entries
belonging to the same block are in red, and the diagonal entries are
in yellow.} \label{fig:Gram}
\end{figure}

Condition \eqref{bound2} is a worst case bound and does not
represent the average recovery ability of block-sparse approximation
methods. It does suggest, however, that in order to improve the
average recovery, all pairs of blocks in $E$ should be as orthogonal
as possible and also all pairs of atoms within each block should be
as orthogonal as possible. Inspired by \cite{sapiro}, rather than
minimizing the inter-block coherence $\mu_B$ and the sub-block
coherence $\nu$, we aim at minimizing the {\em total inter-block
coherence} $\mu_B^t$ and the {\em total sub-block coherence} $\nu^t$
of the equivalent dictionary $E$. We define the total inter-block
coherence as
\begin{equation}
    \label{mu}
    \mu_B^t = \sum_{j=1}^B\sum_{i\neq j}\|G[i,j]\|_F^2,
\end{equation}
and the total sub-block coherence by
\begin{equation}
    \label{nu}
    \nu^t = \sum_{j=1}^B\|G[j,j]\|_F^2-\sum_{m=1}^K(G^m_m)^2,
\end{equation}
where $G^m_m$ are the diagonal entries of $G$. The total inter-block
coherence $\mu_B^t$ equals the sum of the squared entries in $G$
belonging to different blocks (the green entries in
Fig.~\ref{fig:Gram}). Since this is the sum of Frobenius norms,
$\mu_B^t$ also equals the sum of the squared singular values of the
cross-correlation blocks in $G$. When $E$ is normalized, $\mu_B^t$
is equivalent to the sum of the squared cosines of all the principal
angles between all pairs of different blocks. The total sub-block
coherence $\nu^t$ measures the sum of the squared off-diagonal
entries belonging to the same block (the red entries in
Fig.~\ref{fig:Gram}). When $E$ is normalized, $\nu^t$ equals the sum
of the squared cosines of all the angles between atoms within the
same block. Note that when the size of the blocks equals one, we get
$\nu^t=0$.

Alternatively, one could define the total inter-block coherence as
the sum of the squared spectral norms (i.e., the largest singular
values) of the cross-correlation blocks in $G$, and the total
sub-block coherence as the sum of the squared maximal off-diagonal
entries of the auto-correlation blocks in $G$. These definitions are
closer to the ones used in condition \eqref{bound2}. The WCM
algorithm presented in the next section can be slightly modified in
order to minimize those measures as well. However, besides the
increased complexity of the algorithm, the results appear to be
inferior compared to minimizing the definitions \eqref{mu} and
\eqref{nu} of $\mu_B^t$ and $\nu^t$. This can be explained by the
fact that maximizing only the smallest principal angle between pairs
of different blocks in $E$ and maximizing the smallest angle between
atoms within the same block, creates a bulk of relatively high
singular values and coherence values. While this may improve the
worst-case bound in \eqref{bound2}, it does not necessarily improve
the average recovery ability of block-sparse approximation methods.

When minimizing the total inter-block coherence and the total
sub-block coherence, we need to verify that the columns of $E$ are
normalized, to avoid the tendency of columns with small norm values
to be underused. Rather than enforcing normalization strongly, we
penalize for columns with norms that deviate from $1$ by defining
the {\em normalization penalty} $\eta$:
\begin{equation}
    \label{eta}
    \eta=\sum_{m=1}^K(G^m_m-1)^2.
\end{equation}
This penalty $\eta$ measures the sum of the squared distances
between the diagonal entries in $G$ (the yellow entries in Fig.
\ref{fig:Gram}) and $1$.

While \cite{sapiro} did not deal with the block-sparse case, it is
straightforward to see that solving \eqref{sap} is equivalent to
minimizing the sum of the normalization penalty, the total
inter-block coherence and the total sub-block coherence:
\begin{align*}
\|E'E-I\|_F^2=&\sum_{j=1}^B\sum_{i\neq
j}\|E[i]'E[j]\|_F^2+\sum_{j=1}^B\|E[j]'E[j]-I\|_F^2\\
=&\sum_{j=1}^B\sum_{i\neq
j}\|G[i,j]\|_F^2+\sum_{j=1}^B\|G[j,j]-I\|_F^2\\
=&\sum_{j=1}^B\sum_{i\neq
j}\|G[i,j]\|_F^2+\sum_{j=1}^B\|G[j,j]\|_F^2\\
&\quad-\sum_{m=1}^K(G^m_m)^2+\sum_{m=1}^K(G^m_m-1)^2\\
=&\eta+\mu_B^t+\nu^t.
\end{align*}
We have shown in the previous section that the objective in
\eqref{sap} is bounded below by $K-M$. Therefore,
\begin{equation}
\eta+\mu_B^t+\nu^t\geq K-M.
\end{equation}
This bound implies a trade-off, and as a consequence, one cannot
minimize $\eta$, $\mu_B^t$ and $\nu^t$ freely. Instead, we propose
designing a sensing matrix that minimizes the normalization penalty
and a weighted sum of the total inter-block coherence and the total
sub-block coherence:
\begin{equation}
    A = \arg\min_A \frac{1}{2}\eta+ (1-\alpha)\mu_B^t + \alpha
\nu^t, \label{opt1}
\end{equation}
where $0<\alpha<1$ is a parameter controlling the weight given to
the total inter-block coherence and the total sub-block coherence.
Note that alternative objectives can be formulated. For example, one
could add an additional weighting parameter to the normalization
penalty term. While this would allow us to better control the
normalization of the atoms in $E$, we prefer to deal with a single
parameter only.

\begin{figure}
\centering \subfigure[]{
\includegraphics[angle=0,scale=0.4]{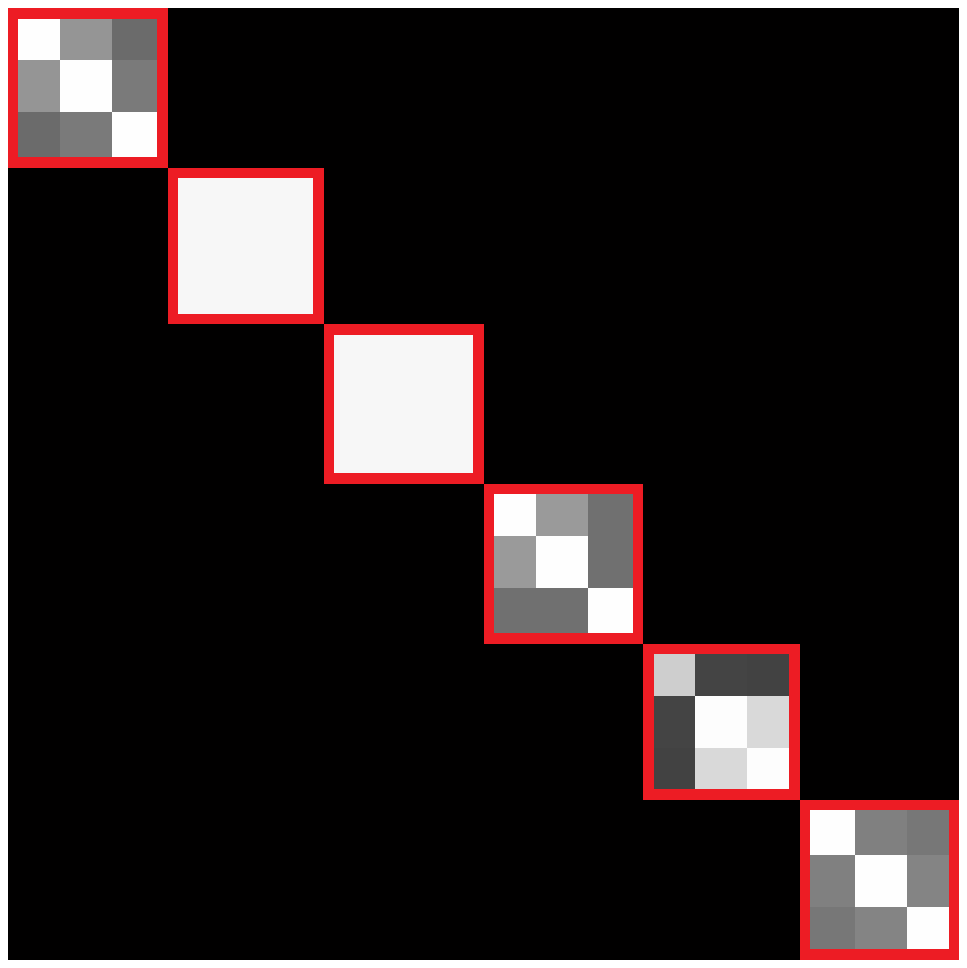}
\label{fig:WCM_low_a} } \hspace{0cm} \subfigure[]{
\includegraphics[angle=0,scale=0.4]{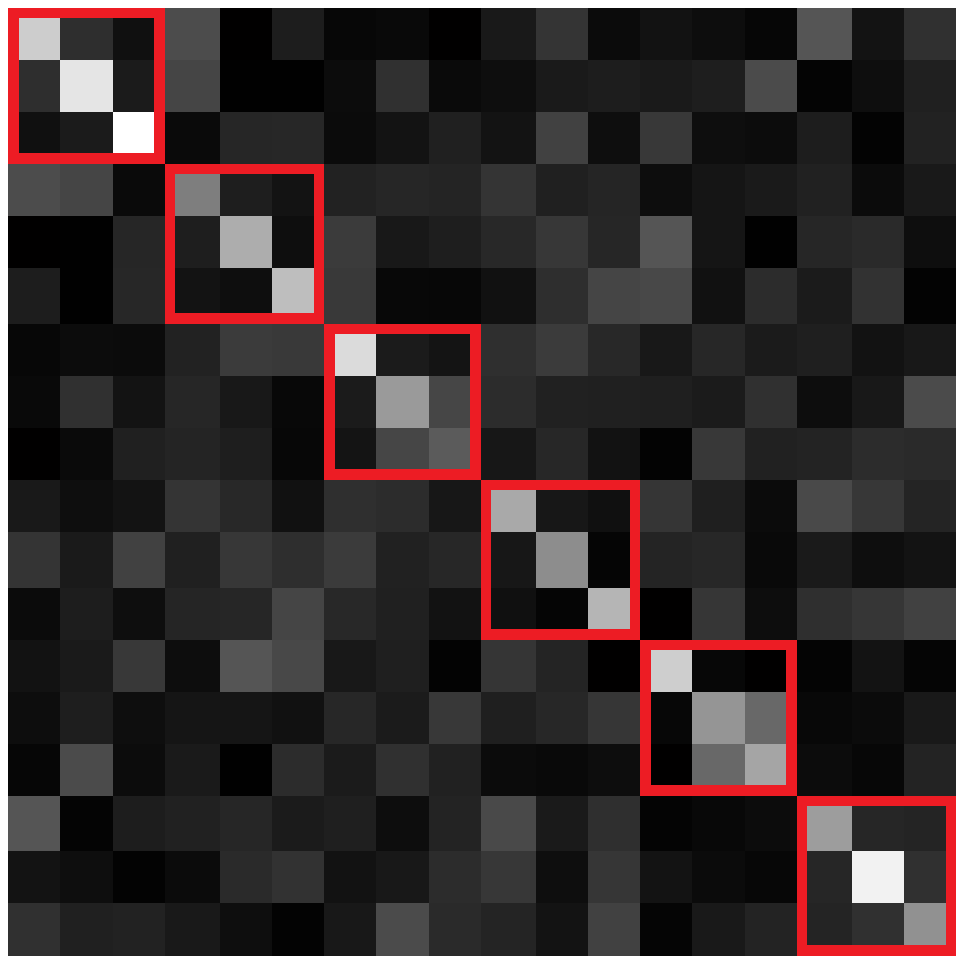}
\label{fig:WCM_medium_a} } \hspace{0cm}\subfigure[]{
\includegraphics[angle=0,scale=0.4]{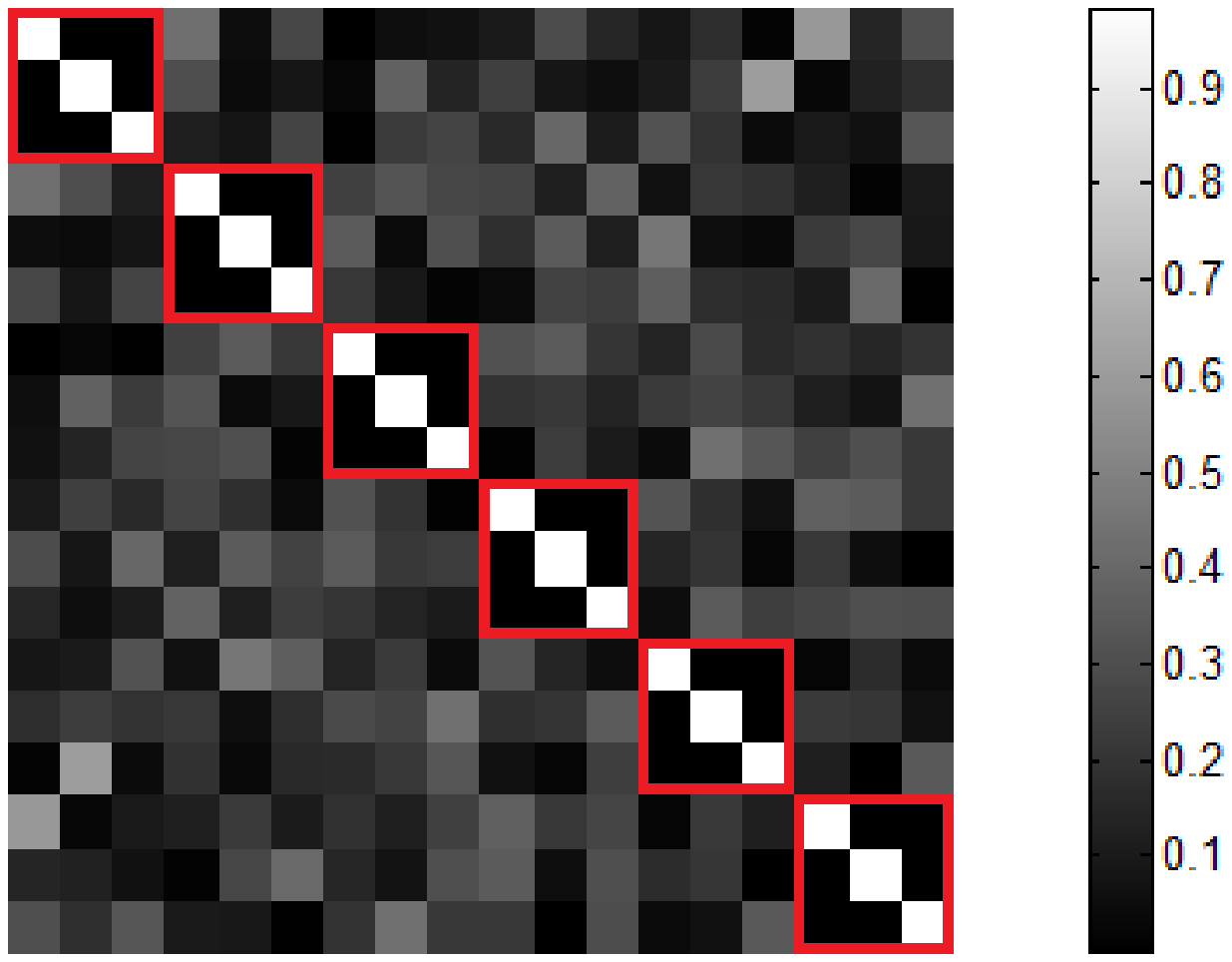}
\label{fig:WCM_high_a} } \hspace{0cm} \caption{Examples of the
absolute value of the Gram matrix of an equivalent dictionary for
$\alpha=0.01$ (a), $\alpha=0.5$ (b) and $\alpha=0.99$ (c), where the
sensing matrix of size $12 \times 18$ was found by solving
\eqref{opt1} given a randomly selected square dictionary composed of
$6$ blocks of size $3$. The sub-block entries are highlighted by red
squares. }
\end{figure}

When $\alpha < \frac{1}{2}$, more weight is given to minimizing
$\mu_B^t$, and therefore solving \eqref{opt1} leads to lower total
inter-block coherence, which is made possible by aligning the atoms
within each block (Fig. \ref{fig:WCM_low_a}). On the other hand,
choosing $\alpha > \frac{1}{2}$ gives more weight to minimizing
$\nu^t$. In this case, solving \eqref{opt1} leads to more
orthonormal blocks in $E$ at the expense of higher $\mu_B^t$ (Fig.
\ref{fig:WCM_high_a}). Finally, setting $\alpha=\frac{1}{2}$ in
\eqref{opt1} gives equal weights to $\mu_B^t$, $\nu^t$ and $\eta$,
and reduces it to \eqref{sap} (Fig. \ref{fig:WCM_medium_a}).
Therefore, the objective becomes independent of the block structure,
which makes $\alpha=\frac{1}{2}$ the correct choice when the signals
do not have an underlying block structure. Choosing to ignore the
block structure leads to the same conclusion. When an underlying
block structure exists, we need to select a value for $\alpha$. We
do that via empirical evaluation in Section~\ref{sec:experiments}.

\begin{figure}
\centering
\includegraphics[width=100mm,angle=0,scale=0.5]{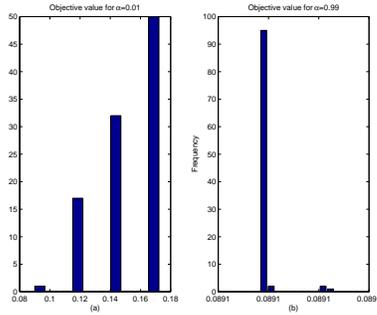}
\caption{Histograms of the objective values obtained when solving
\eqref{opt1} $100$ times with $\alpha=0.01$ (a) and $\alpha=0.99$
(b), for a given randomly generated square dictionary composed of
$6$ blocks of size $3$. The sensing matrices of size $12 \times 18$
are initialized as matrices with random entries. Note that the
distribution is insignificant in (b), indicating that in this
specific case, every local minimum is also a global minimum.}
\label{fig:hist}
\end{figure}

\begin{figure}
\centering
\includegraphics[width=100mm,angle=0,scale=0.5]{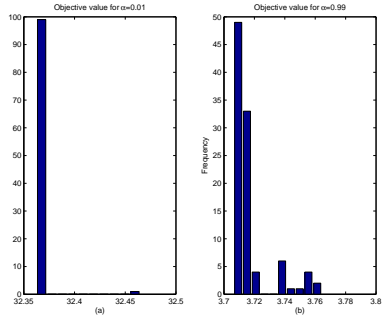}
\caption{Histograms of the objective values obtained when solving
\eqref{opt1} $100$ times with $\alpha=0.01$ (a) and $\alpha=0.99$
(b), for a given randomly generated overcomplete dictionary composed
of $24$ blocks of size $3$. The sensing matrices of size $12 \times
18$ are initialized as matrices with random entries.}
\label{fig:hist4N}
\end{figure}

In the previous section we have shown that every local minimum of
\eqref{sap}, and therefore also of \eqref{opt1} with
$\alpha=\frac{1}{2}$, is also a global minimum. Empirical
observations reveal that this is not the case when $\alpha \neq
\frac{1}{2}$. This is demonstrated in the histograms presented in
Fig.~\ref{fig:hist}(a) for $\alpha = 0.01$ with a square dictionary
and in Fig.~\ref{fig:hist4N}(b) for $\alpha = 0.99$ with a highly
overcomplete dictionary. Since it is hard to develop a closed form
solution for \eqref{opt1}, we present an iterative algorithm that
converges to a local solution of \eqref{opt1} in the following
section.

\section{Weighted Coherence Minimization}
\label{sec:WCM}

In this section, we present the \emph{Weighted Coherence
Minimization} (WCM) algorithm for minimizing \eqref{opt1}, based on
the bound-optimization method \cite{BO}. This algorithm substitutes
the original objective with an easier to minimize surrogate
objective that is updated in each optimization step. After defining
a surrogate function and showing it can be minimized in closed form,
we prove that its iterative minimization is guaranteed to converge
to a local solution of the original problem.

\subsection{The Weighted Coherence Minimization Algorithm}
\label{ssec:alg}

To obtain a surrogate function we rewrite the objective of
\eqref{opt1}, which we denote by $f(G)$, as a function of the Gram
matrix of the equivalent dictionary $G = D'A'AD$:
\begin{align*}
f(G)\equiv&\frac{1}{2}\eta(G)+(1-\alpha)\mu_B^t(G) + \alpha \nu^t(G)\\
    =&\frac{1}{2}\|u_\eta(G)\|_F^2+(1-\alpha)\|u_\mu(G)\|_F^2+\alpha\|u_\nu(G)\|_F^2,
\end{align*}
where the matrix operators $u_\mu$, $u_\nu$ and $u_\eta$ are defined
as:
\begin{align*}
    u_\eta(G)[i,j]^m_n &= \left\{
                         \begin{array}{ll}
                           G[i,j]^m_n-1, & \hbox{$i=j, m = n$;} \\
                           0, & \hbox{else}
                         \end{array}\right.,\\
    u_\mu(G)[i,j]^m_n &= \left\{
                         \begin{array}{ll}
                           G[i,j]^m_n, & \hbox{$i\neq j$;} \\
                           0, & \hbox{else}
                         \end{array}\right.,\\
    u_\nu(G)[i,j]^m_n &= \left\{
                         \begin{array}{ll}
                           G[i,j]^m_n, & \hbox{$i=j, m \neq n$;} \\
                           0, & \hbox{else}
                         \end{array}\right.,\\
\end{align*}
with $G[i,j]^m_n$ denoting the $(m,n)$th entry of $G[i,j]$. This
equation follows directly from the definitions of $\eta$, $\mu_B^t$
and $\nu^t$. We can now write:
\begin{align}
   f(G) = \frac{1}{2}\|G-h_\eta(G)\|_F^2+(1-\alpha)&\|G-h_\mu(G)\|_F^2\nonumber\\
 +\alpha&\|G-h_\nu(G)\|_F^2,
\label{molding_obj}
\end{align}
where the matrix operators $h_\mu$, $h_\nu$ and $h_\eta$ are defined
as:
\begin{align*}
    h_\eta(G)[i,j]^m_n &= \left\{
                         \begin{array}{ll}
                           1, & \hbox{$i=j, m = n$;} \\
                           G[i,j]^m_n, & \hbox{else}
                         \end{array}\right.,\\
    h_\mu(G)[i,j]^m_n &= \left\{
                         \begin{array}{ll}
                           0, & \hbox{$i\neq j$;} \\
                           G[i,j]^m_n, & \hbox{else}
                         \end{array}\right.,\\
    h_\nu(G)[i,j]^m_n &= \left\{
                         \begin{array}{ll}
                           0, & \hbox{$i=j, m \neq n$;} \\
                           G[i,j]^m_n, & \hbox{else}
                         \end{array}\right..
\end{align*}

Based on \eqref{molding_obj}, we define a surrogate objective
$g(G,G^{(n)})$ at the $n$th iteration as:
\begin{align}
g(G,G^{(n)})\equiv\frac{1}{2}\|G-h_\eta(G^{(n)})\|_F^2+(1-\alpha)&\|G-h_\mu(G^{(n)})\|_F^2\nonumber\\
+\alpha&\|G-h_\nu(G^{(n)})\|_F^2, \label{molding}
\end{align}
where $G^{(n)} = D'A^{(n)'}A^{(n)}D$ is the Gram matrix of the
equivalent dictionary from the previous iteration. In Appendix B, we
prove that $g(G,G^{(n)})$ satisfies the conditions of a surrogate
objective for the bound-optimization method. Therefore, iteratively
minimizing $g(G,G^{(n)})$ is guaranteed to converge to the minimum
of the original objective $f(G)$, i.e., solve \eqref{opt1}.

The following proposition describes the closed form solution to
minimizing $g(G,G^{(n)})$ at each iteration.

\textbf{Proposition $1$:} The function $g(G,G^{(n)})$ is minimized
by choosing
\[A^{(n+1)} = \Delta_M^{1/2}V_M'\Lambda^{-1/2}U',\] where $U\Lambda
U'$ is the eigenvalue decomposition of $DD'$, $\Delta_M$ and $V_M$
are the top $M$ eigenvalues and the corresponding $M$ eigenvectors
of $\Lambda^{-1/2}U'Dh_t(G^{(n)})D'U\Lambda^{-1/2}$, and:
\begin{align}
h_t(\cdot)\equiv\frac{2}{3}\left(\frac{1}{2}h_\eta(\cdot)+(1-\alpha)h_\mu(\cdot)+\alpha
h_\nu(\cdot)\right).\label{h_t}
\end{align}

\begin{proof}See Appendix B.
\end{proof}

A summary of the proposed WCM algorithm is given below.

\begin{algorithm}
\caption{Weighted Coherence Minimization}
 \emph{Task}: Solve for a given block-sparsifying dictionary $D_{N \times K}$:
\begin{equation*}
    A = \arg\min_A \frac{1}{2}\eta+ (1-\alpha)\mu_B^t + \alpha
\nu^t,
\end{equation*}
where $A \in R^{M\times N}$. \newline
\emph{Initialization}:
Calculate the eigenvalue decomposition of $DD'=U\Lambda U'$. Set
$A^{(0)}$ as the outcome of \eqref{sap}, i.e.,
$A^{(0)} = [I_M \textrm{ } 0] \Lambda^{-1/2} U'$, and $n=0$.\\
\emph{Repeat until convergence}:
\begin{enumerate}
  \item Set $G^{(n)}=D'A^{(n)'}A^{(n)}D$.
  \item Calculate $h_t(G^{(n)})$ as in \eqref{h_t}.
  \item Find the top $M$ eigenvalues $\Delta_M$ and the corresponding $M$
        eigenvectors $V_M$ of
        $\Lambda^{-1/2}U'Dh_t(G^{(n)})D'U\Lambda^{-1/2}$.
  \item Set $A^{(n+1)} = \Delta_M^{1/2}V_M'\Lambda^{-1/2}U'$.
  \item $n=n+1$.
\end{enumerate}
\end{algorithm}

\section{Experiments}
\label{sec:experiments}

In this section, we evaluate the contribution of the proposed
sensing matrix design framework empirically. We compare the recovery
and classification abilities of BOMP \cite{BOMP1,BOMP2} when using
sensing matrices designed by our methods to the outcome of
\eqref{sap}, which will be referred to as ``Duarte-Sapiro'' (DS)
\cite{sapiro}.

For each simulation, we repeat the following procedure $100$ times.
We randomly generate a dictionary $D_{N \times K}$ with normally
distributed entries and normalize its columns. In order to evaluate
WCM on structured dictionaries as well, we repeat the simulations
using a dictionary containing $N$ randomly selected rows of the $K
\times K$ Discrete Cosine Transform (DCT) matrix. The dictionary is
divided into $K/s$ blocks of size $s$. We then generate $L=1000$
test signals $X$ of dimension $K$ that have $k$-block-sparse
representations $\Theta$ with respect to $D$. The generating blocks
are chosen randomly and independently and the coefficients are
i.i.d. uniformly distributed. $A_{M\times N}$ is initialized as the
outcome of DS. We find $A$ using the WCM algorithm, and calculate
the equivalent dictionary $E=AD$ and the measurements $Y=AX$. Next,
we obtain the block-sparsest representations of the measurements,
$\hat{\Theta}$, by applying BOMP with a fixed number of $k$ nonzero
blocks.

We use two measures to evaluate the success of the simulations based
on their outputs $A$ and $\hat{\Theta}$:
\begin{itemize}
  \item The percentage of recognized generating subspaces
  of $X$ (i.e., successful classification): $r=\frac{\|\hat{\Theta}\odot\Theta\|_{0}}{Lks}$
  $\\$ where $\odot$ denotes element-wise multiplication.
  \item The normalized representation error $e=\frac{\|X-D\hat{\Theta}\|_F}{\|X\|_F}$
\end{itemize}

\begin{figure}
\center\centering
\subfigure[]{\includegraphics[width=4.25cm]{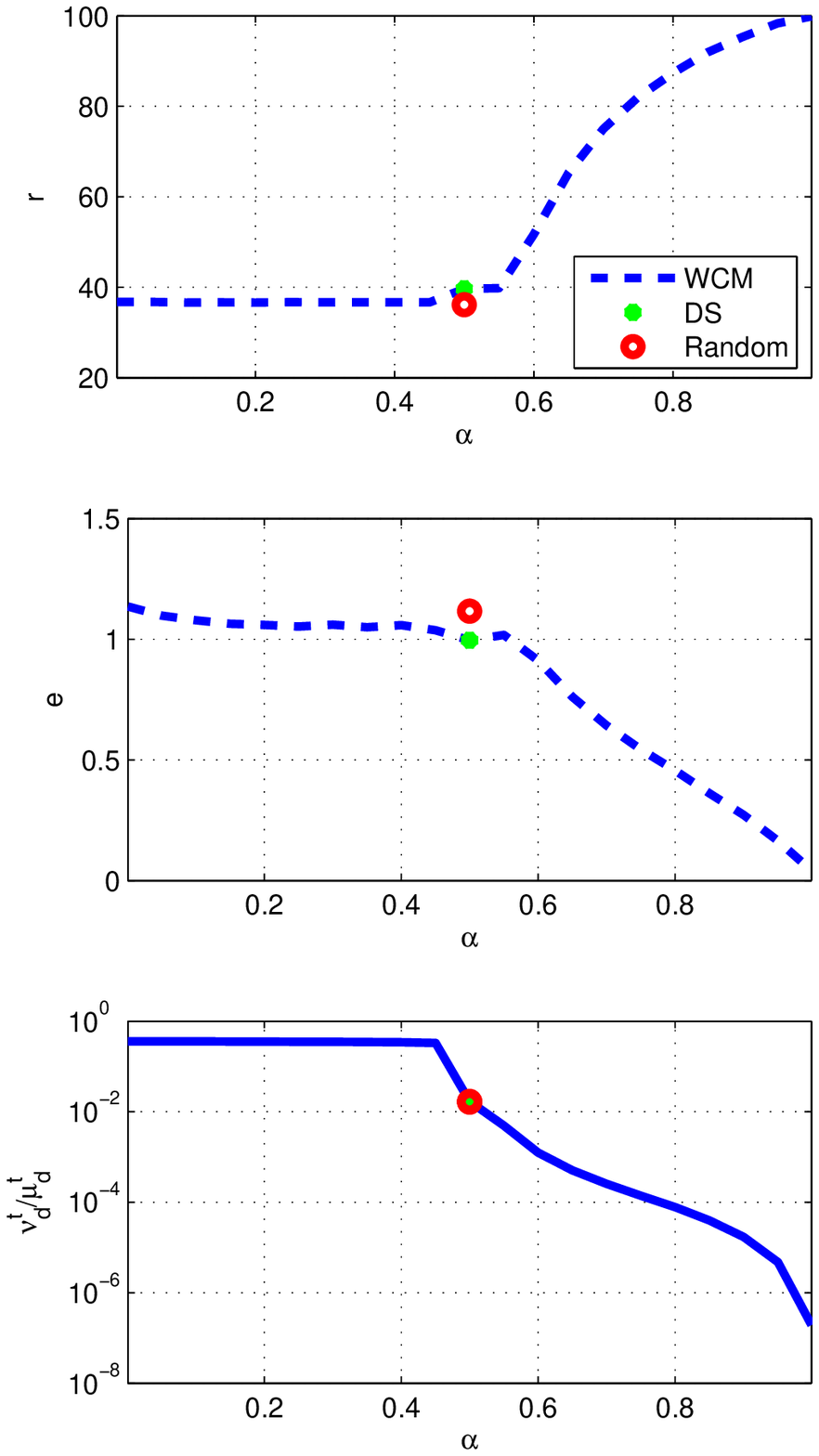}
 \label{fig:WCM_k1}}\hspace{0cm}
\subfigure[]{\includegraphics[width=4.25cm]{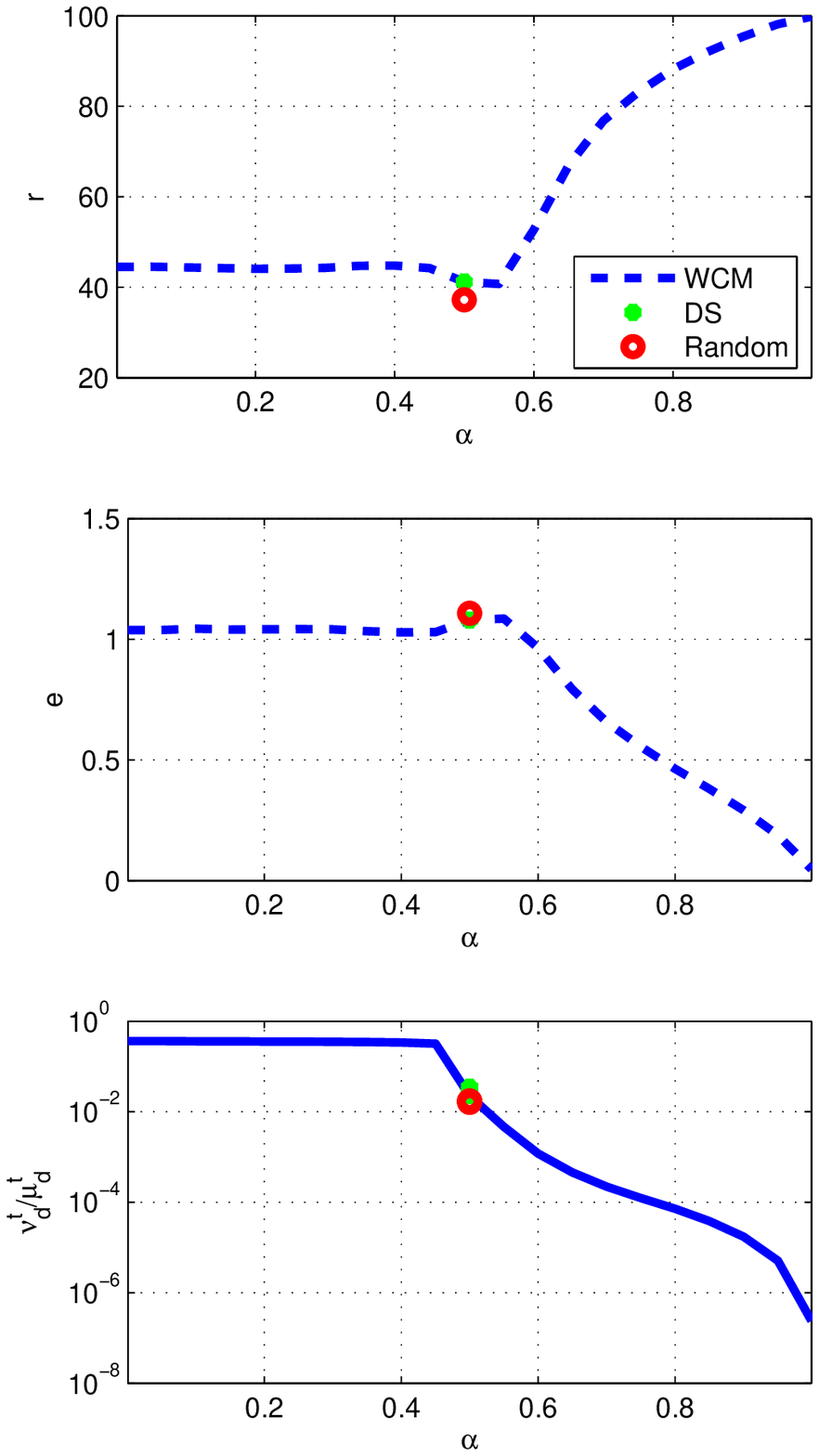}
 \label{fig:WCM_k1_DCT}}
\caption{Simulation results of sensing matrix design using the WCM
algorithm with $k=1$ and $M=6$. The graphs show the normalized
representation error $e$, the classification success $r$, and the
ratio between the total sub-block coherence and the total
inter-block coherence $\nu^t/\mu_B^t$ as a function of $\alpha$. In
(a) the dictionary contains normally distributed entries, and in (b)
randomly selected rows of the DCT matrix.}
\end{figure}

\begin{figure}
\center\centering
\subfigure[]{\includegraphics[width=4.25cm]{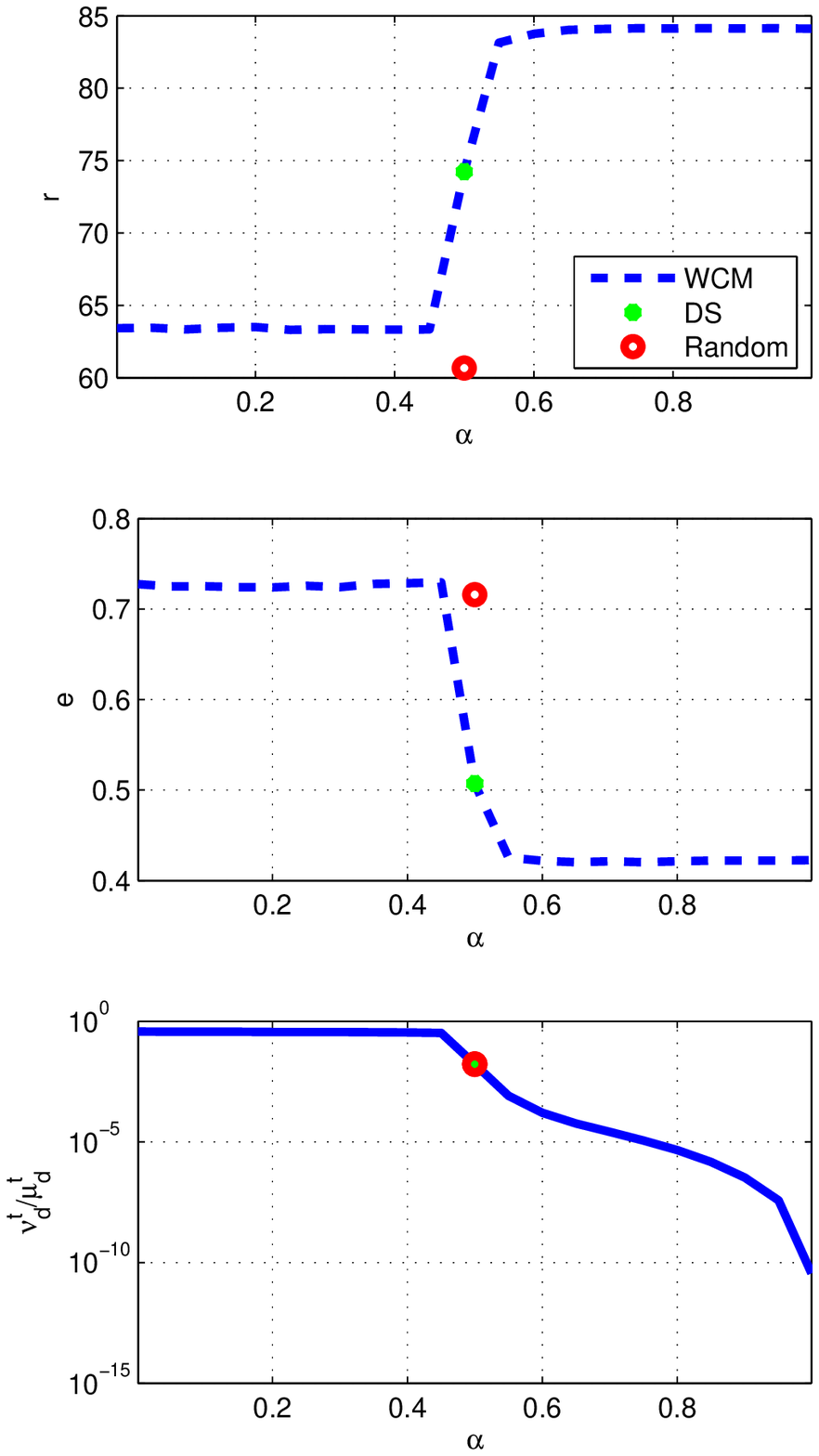}
 \label{fig:WCM_k2}}\hspace{0cm}
\subfigure[]{\includegraphics[width=4.25cm]{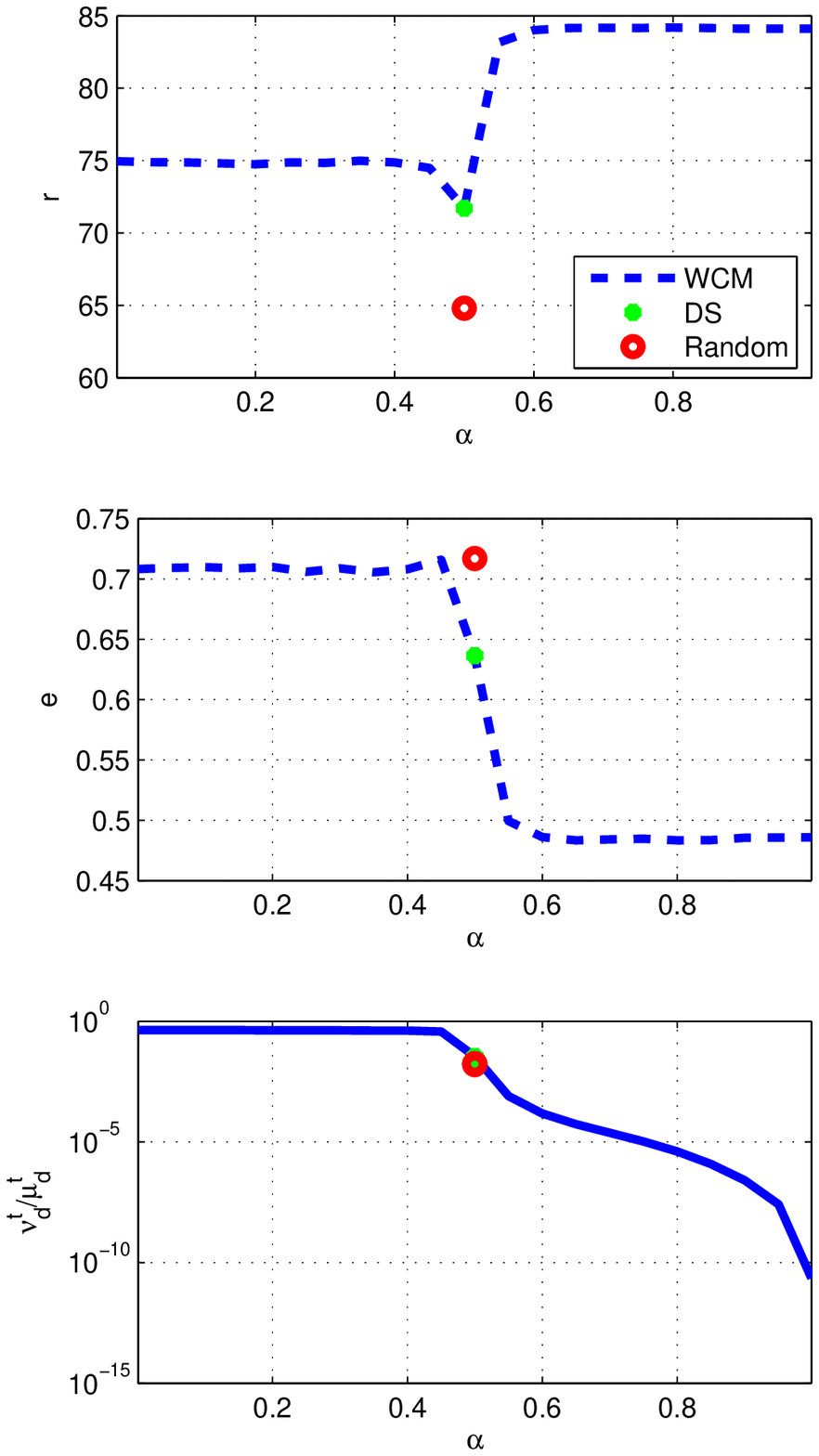}
 \label{fig:WCM_k2_DCT}}
\caption{Simulation results of sensing matrix design using the WCM
algorithm with $k=2$ and $M=14$. The graphs show the normalized
representation error $e$, the classification success $r$, and the
ratio between the total sub-block coherence and the total
inter-block coherence $\nu^t/\mu_B^t$ as a function of $\alpha$. In
(a) the dictionary contains normally distributed entries, and in (b)
randomly selected rows of the DCT matrix.}
\end{figure}

\begin{figure}
\center\centering
\subfigure[]{\includegraphics[width=4.25cm]{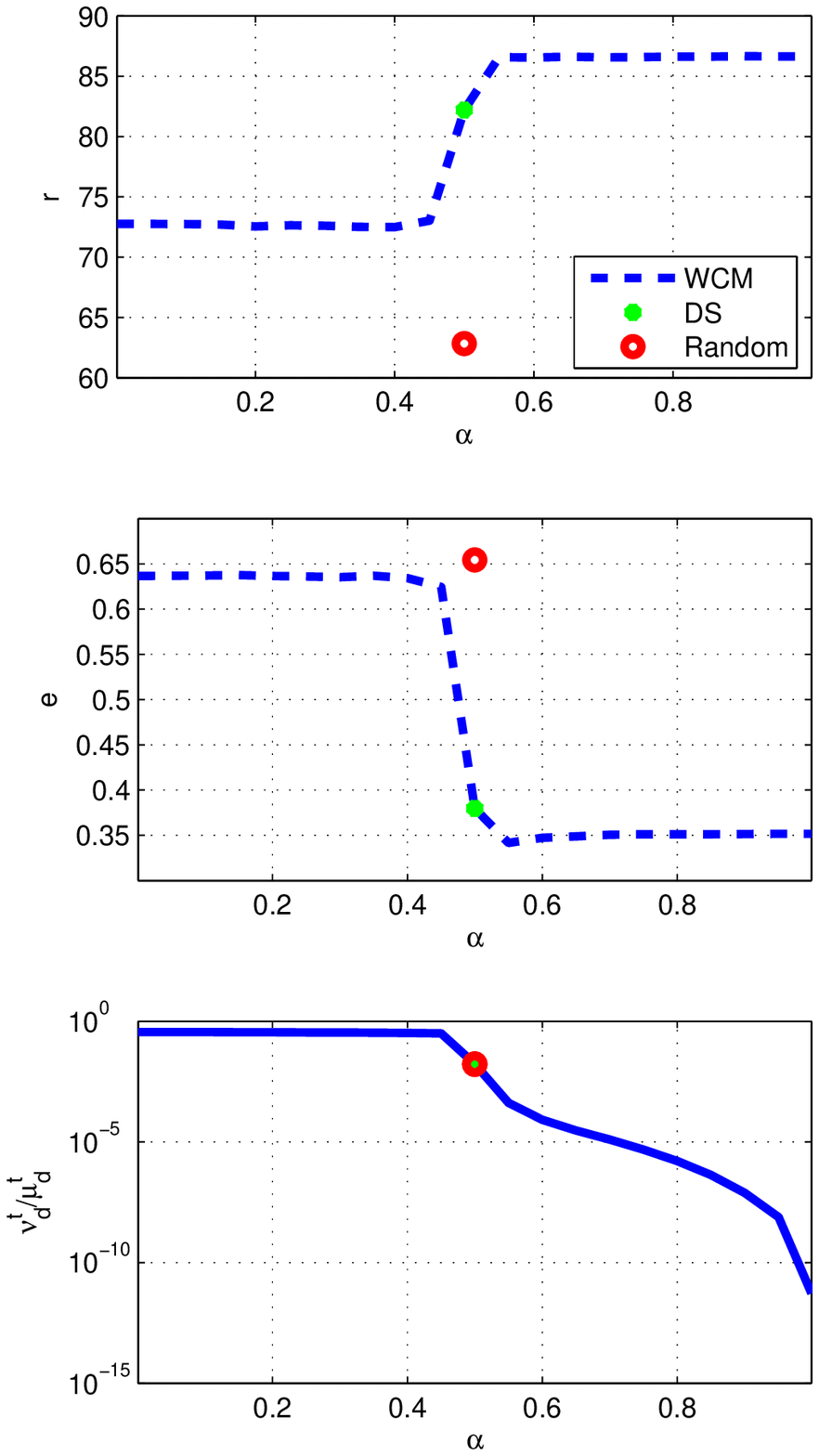}
 \label{fig:WCM_k3}}\hspace{0cm}
\subfigure[]{\includegraphics[width=4.25cm]{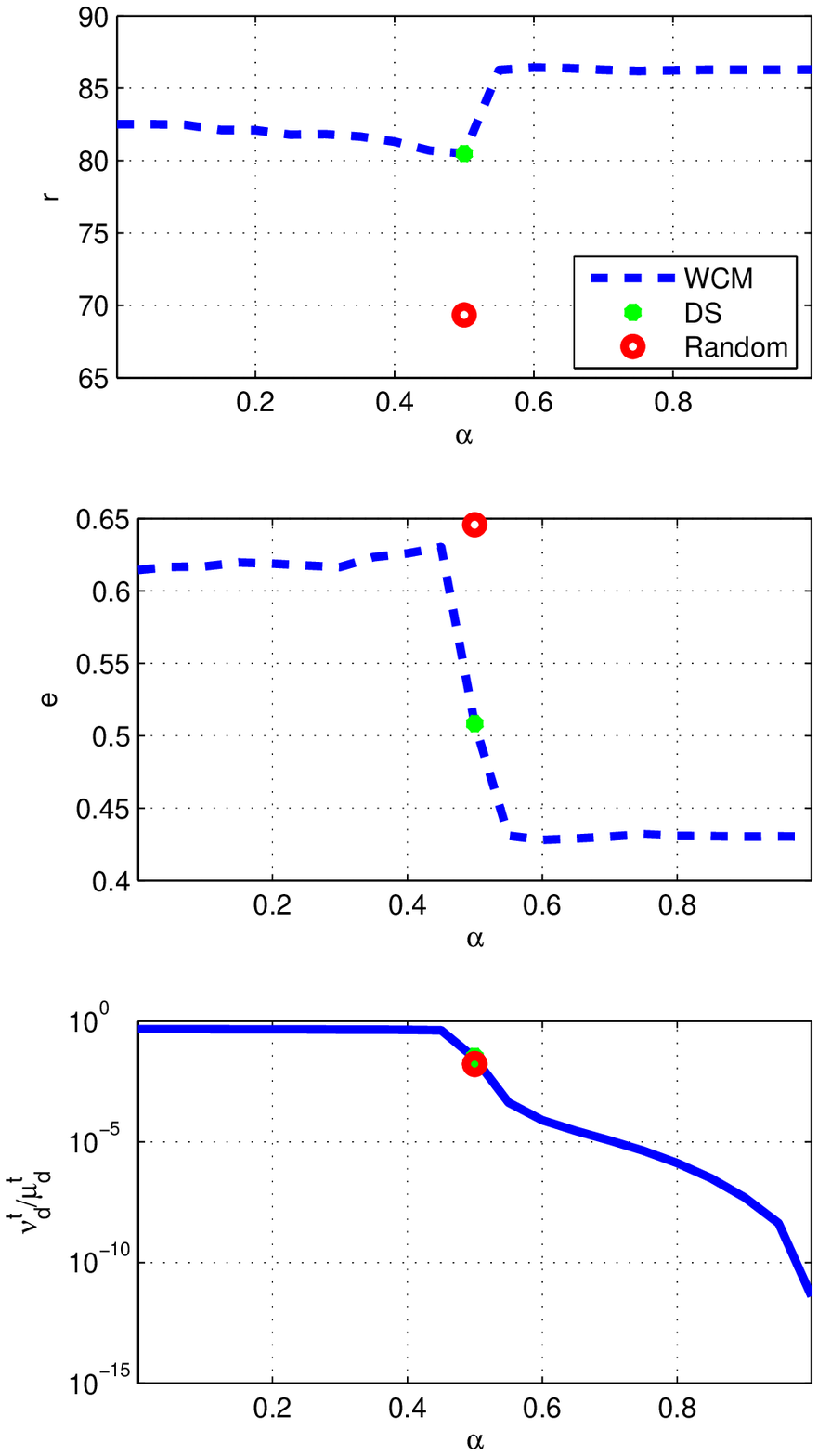}
 \label{fig:WCM_k3_DCT}}
\caption{Simulation results of sensing matrix design using the WCM
algorithm with $k=3$ and $M=20$. The graphs show the normalized
representation error $e$, the classification success $r$, and the
ratio between the total sub-block coherence and the total
inter-block coherence $\nu^t/\mu_B^t$ as a function of $\alpha$. In
(a) the dictionary contains normally distributed entries, and in (b)
randomly selected rows of the DCT matrix.}
\end{figure}

\begin{figure}
\center\centering
\subfigure[]{\includegraphics[width=4.25cm]{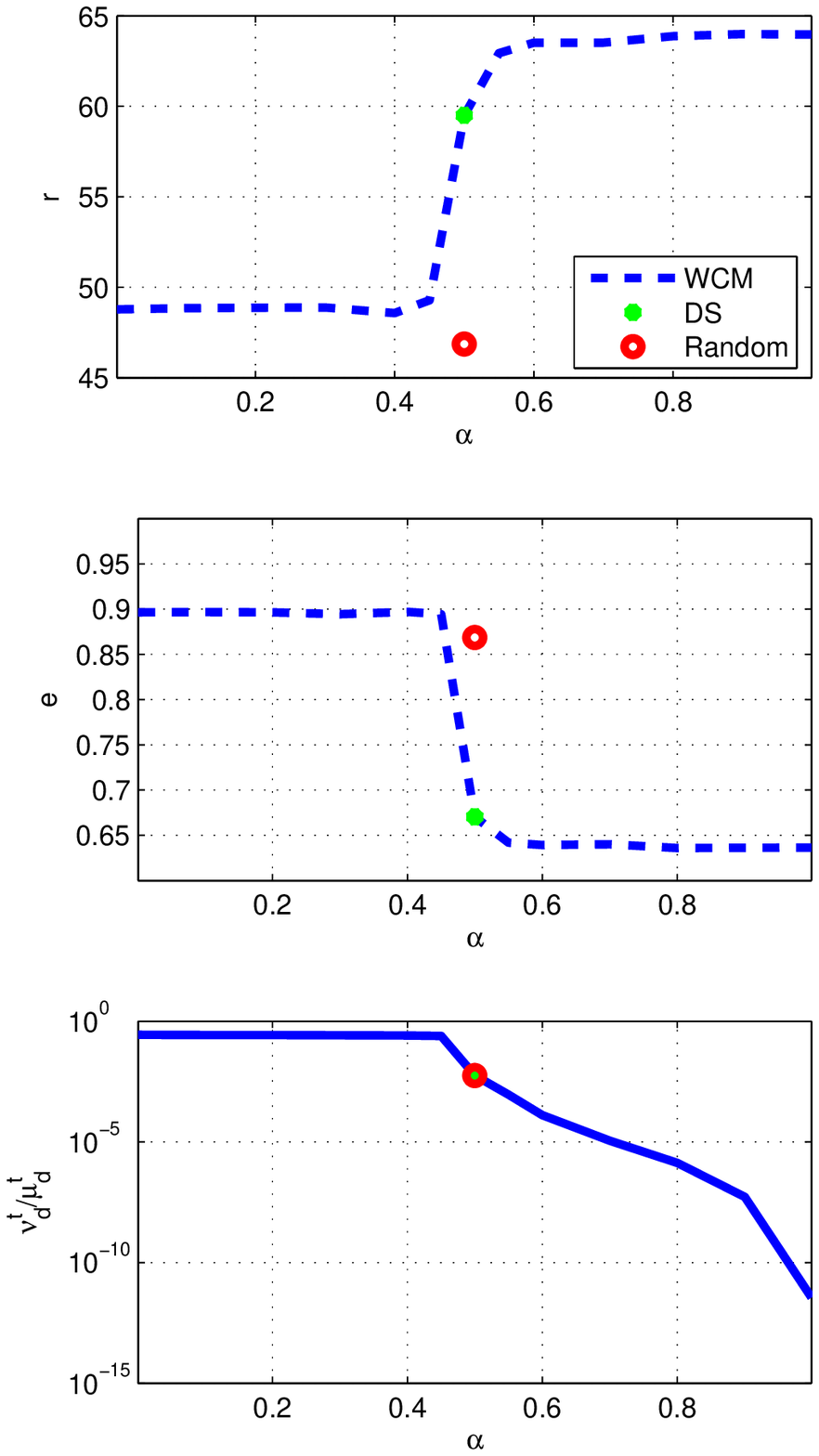}
 \label{fig:WCM_k6}}\hspace{0cm}
\subfigure[]{\includegraphics[width=4.25cm]{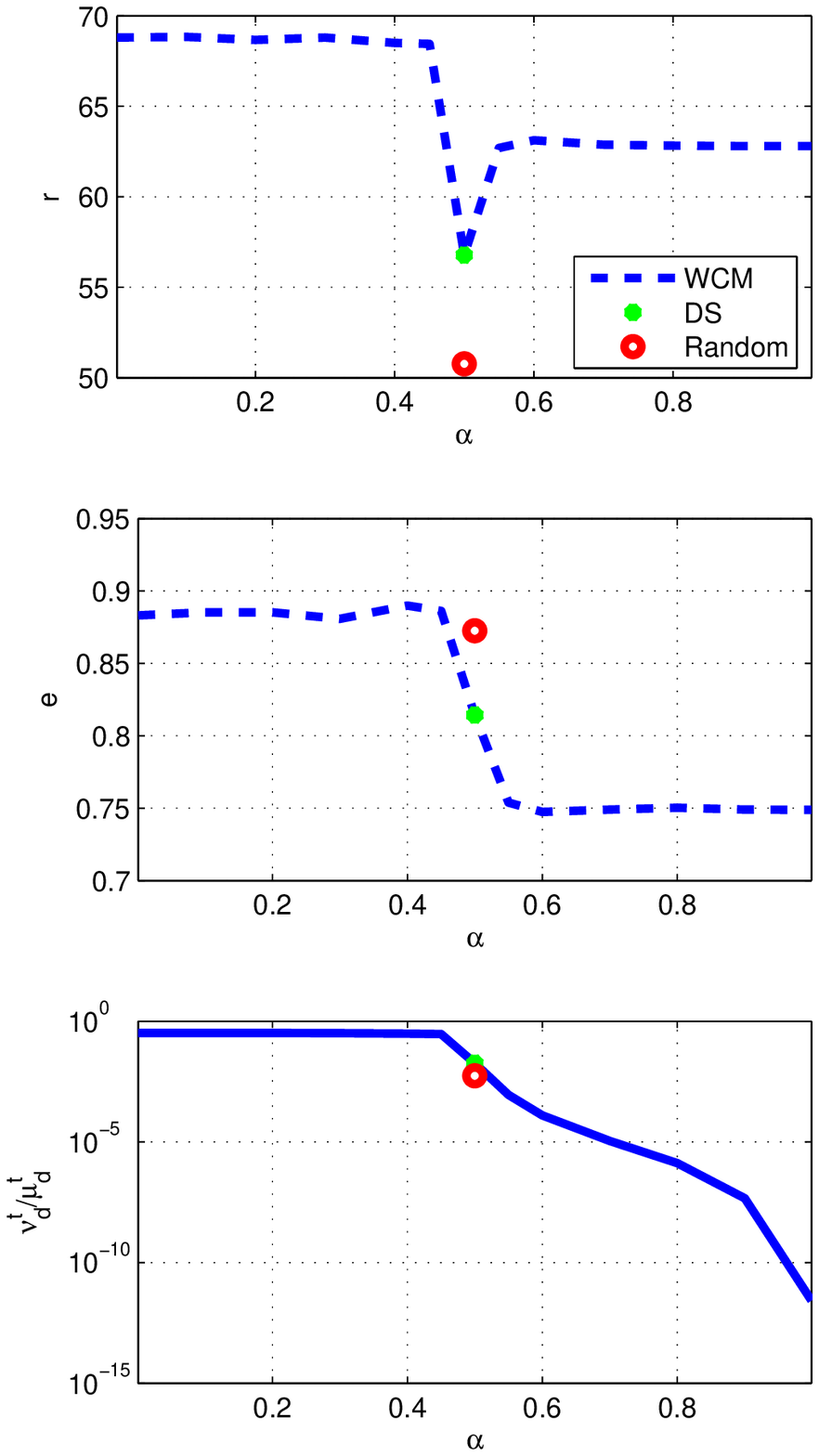}
 \label{fig:WCM_k6_DCT}}
\caption{Simulation results of sensing matrix design using the WCM
algorithm with $k=6$ and $M=35$. The graphs show the normalized
representation error $e$, the classification success $r$, and the
ratio between the total sub-block coherence and the total
inter-block coherence $\nu^t/\mu_B^t$ as a function of $\alpha$. In
(a) the dictionary contains normally distributed entries, and in (b)
randomly selected rows of the DCT matrix.}
\end{figure}

To evaluate the performance of the WCM algorithm as a function of
$\alpha$, we choose $s=3$, $N=60$ and $K=2N=120$. We repeat the
experiment for both types of dictionaries, and for $k=1$
(Fig.~\ref{fig:WCM_k1},\ref{fig:WCM_k1_DCT}), $k=2$
(Fig.~\ref{fig:WCM_k2},\ref{fig:WCM_k2_DCT}) and $k=3$
(Fig.~\ref{fig:WCM_k3},\ref{fig:WCM_k3_DCT}) nonzero blocks, with
respectively $M=6$, $M=14$ and $M=20$ measurements. To show that the
results remain consistent for higher values of $k$, we add an
experiment with $k=6$, $M=35$, $N=180$ and $K=2N=360$
(Fig.~\ref{fig:WCM_k6},\ref{fig:WCM_k6_DCT}). We compare the
obtained results to randomly set sensing matrices and to the outputs
of DS \cite{sapiro}, based on the normalized representation error
$e$, the classification success $r$, and the ratio between the total
sub-block coherence and the total inter-block coherence
$\nu^t/\mu_B^t$. We observe that WCM and DS coincide at $\alpha=0.5$
for all the three measures, as expected. Note that for $\alpha<0.5$
we get that $\nu^t/\mu_B^t$ is high, $e$ is high and $r$ is low. On
the other hand, when $\alpha>0.5$, i.e., when giving more weight to
$\nu^t$ and less to $\mu_B^t$, the signal reconstruction as well as
the signal classification are improved compared to DS. While the
improvement for $k=1$ is more significant, it is maintained for
higher values of $k$ as well. Remarkably, for structured
dictionaries and for higher values of $k$, we see that $\alpha<0.5$
leads to an improvement of $r$. However, $e$ is compromised in this
case. We can conclude that when designing sensing matrices for block
sparse decoding, the best results are obtained by choosing $\alpha$
close enough to $1$. In other words, the best recovery results are
obtained when the equivalent dictionary has nearly orthonormal
blocks. This holds for dictionaries containing normally distributed
entries as well as for dictionaries containing randomly selected
rows of the DCT matrix. As was the case in Fig.~\ref{fig:hist}(b),
we observed empirically that for $\alpha>0.5$, every local minimum
is a global minimum as well. This means that the WCM algorithm
converges to a global solution of \eqref{opt1} when $\alpha>0.5$,
for all the experiments presented in this section. We emphasize
however, that this may not be the case for other sets of parameters.

\begin{figure}
\center\centering
\subfigure[]{\includegraphics[width=4.25cm]{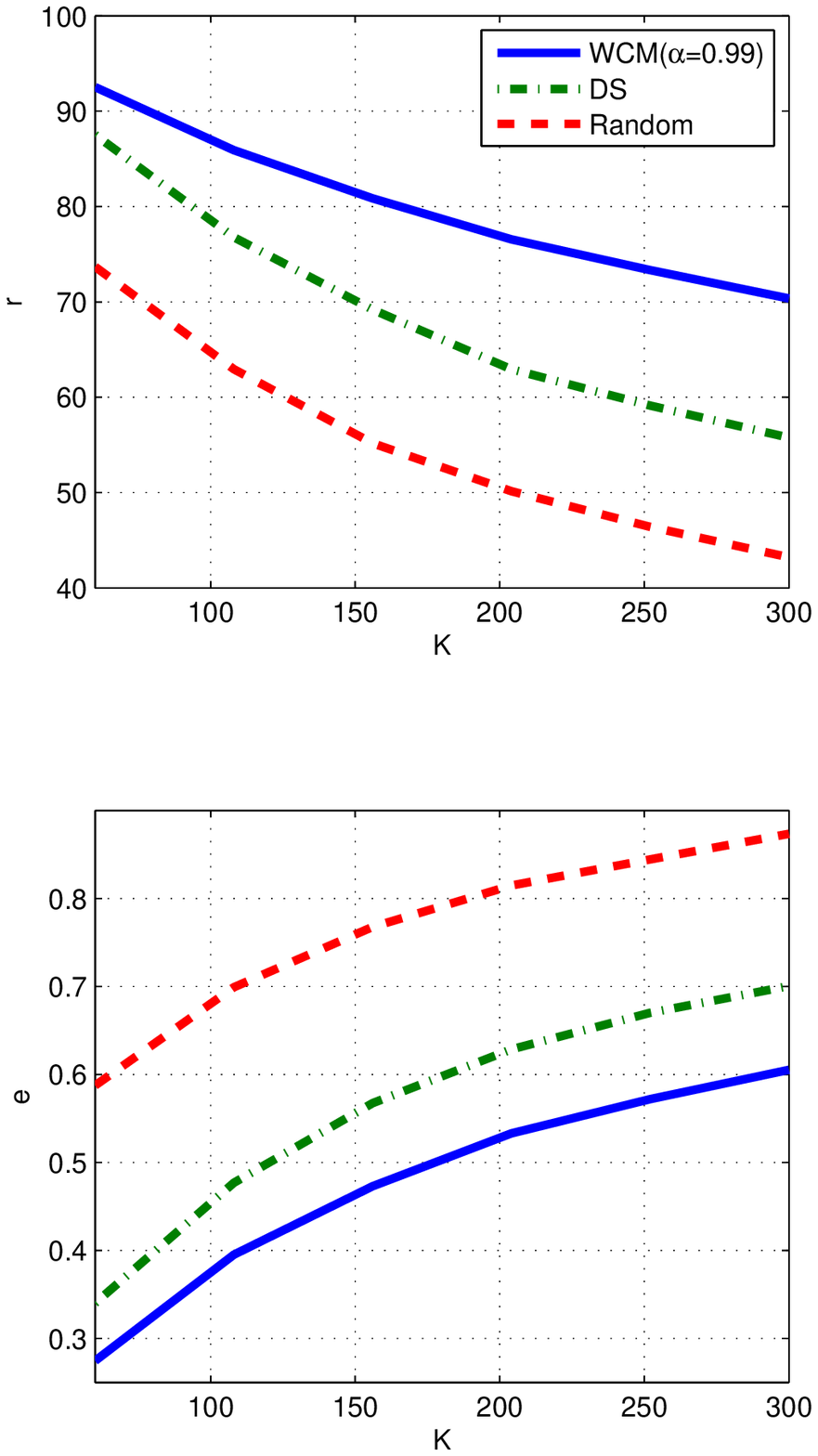}
 \label{fig:WCM_K}}\hspace{0cm}
\subfigure[]{\includegraphics[width=4.25cm]{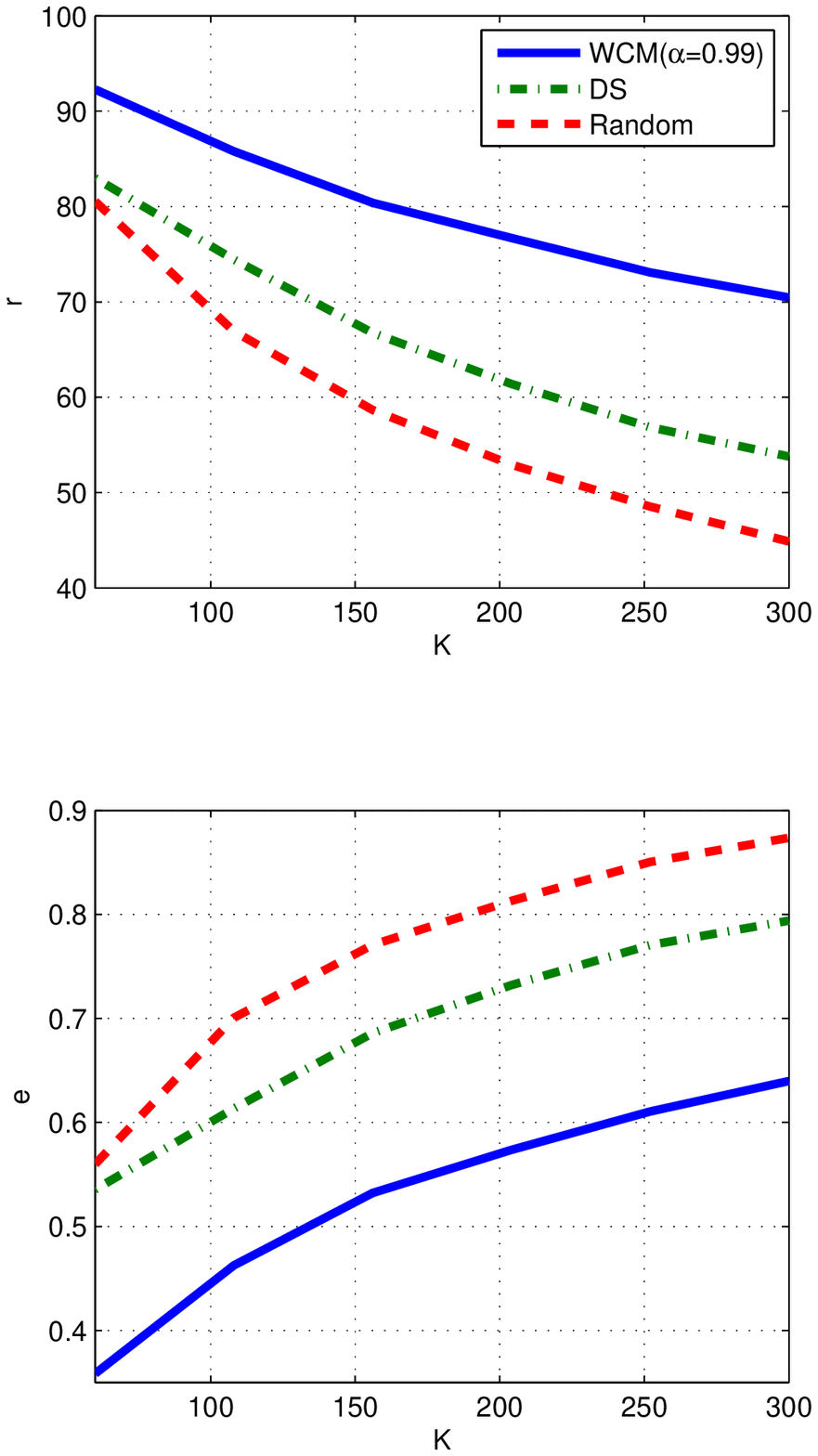}
 \label{fig:WCM_K_DCT}}
\caption{Simulation results of sensing matrix design using the WCM
algorithm with $k=2$ and $M=14$. The graphs show the normalized
representation error $e$ and the classification success $r$ as a
function of $K$. In (a) the dictionary contains normally distributed
entries, and in (b) randomly selected rows of the DCT matrix.}
\end{figure}

Fig.~\ref{fig:WCM_K} and Fig.~\ref{fig:WCM_K_DCT} show that when
using WCM with $\alpha = 0.99$ on dictionaries with normally
distributed entries and on structured dictionaries, the improvement
in signal recovery using is maintained for a wide range of $K$,
starting from square dictionaries, i.e. $K=N$, to highly
overcomplete dictionaries. For this experiment, we chose $s=3$,
$N=60$, $k=2$ and $M=14$. We note that for both types of
dictionaries, the improvement of WCM over DS increases as the
dictionary becomes more overcomplete.

\begin{figure}
\center\centering
\subfigure[]{\includegraphics[width=4.25cm]{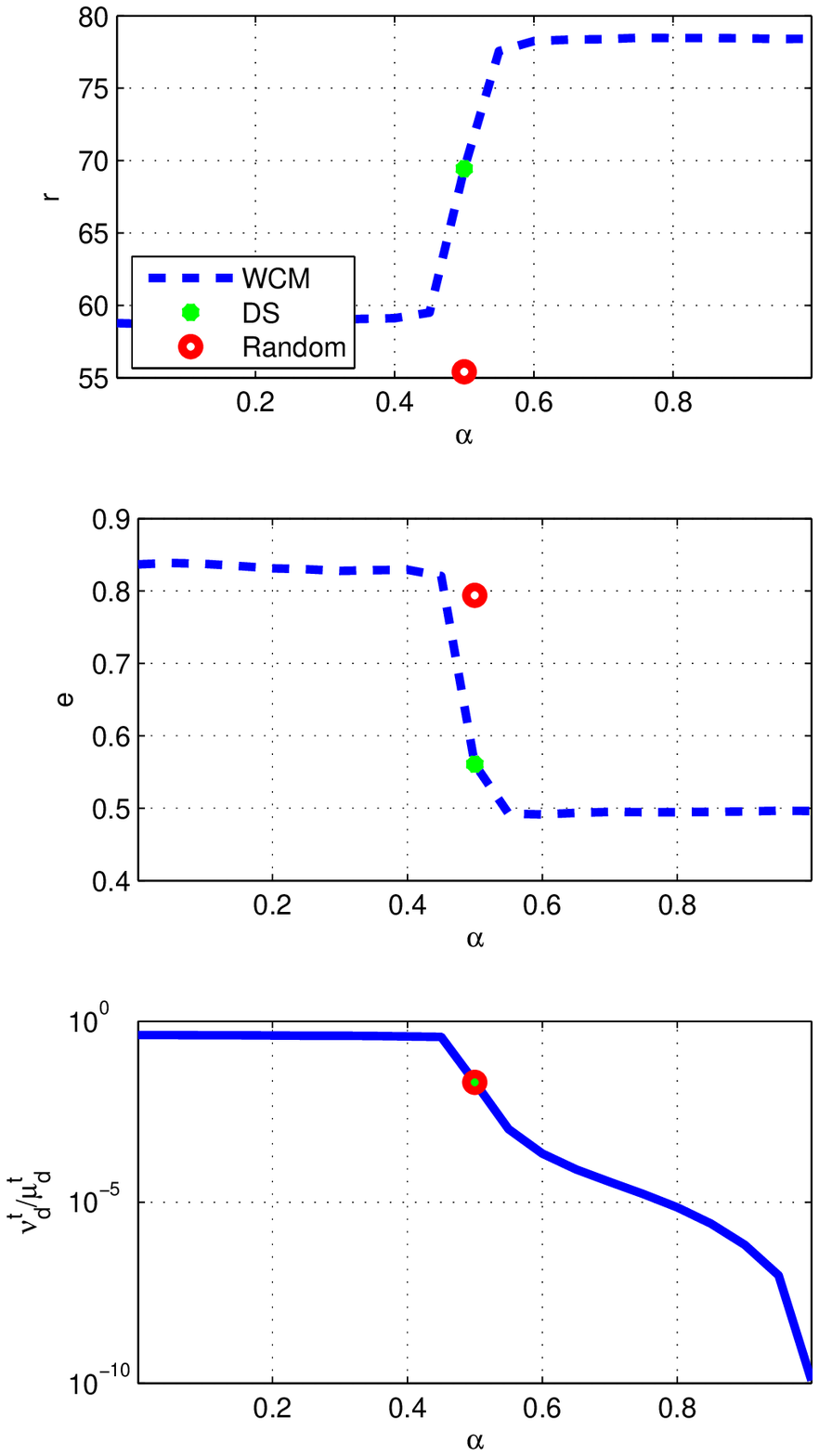}
 \label{fig:WCM_s}}\hspace{0cm}
\subfigure[]{\includegraphics[width=4.25cm]{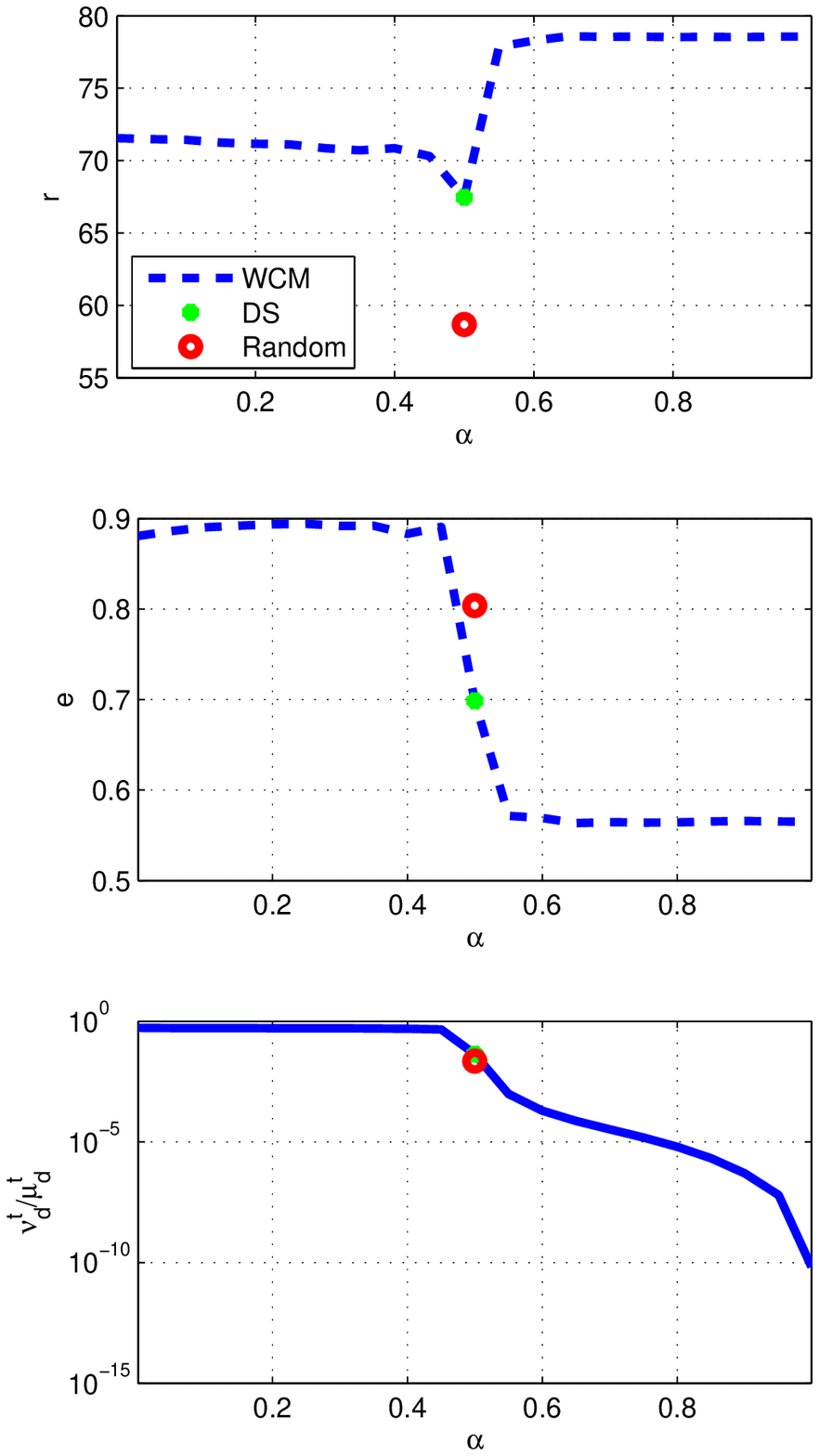}
 \label{fig:WCM_s_DCT}}
\caption{Simulation results of sensing matrix design using the WCM
algorithm on a dictionary containing $15$ blocks of size $4$ and
$20$ blocks of size $3$, with $k=2$ and $M=14$. The graphs show the
normalized representation error $e$, the classification success $r$,
and the ratio between the total sub-block coherence and the total
inter-block coherence $\nu^t/\mu_B^t$ as a function of $\alpha$. In
(a) the dictionary contains normally distributed entries, and in (b)
randomly selected rows of the DCT matrix.}
\end{figure}

Finally, we show that WCM improves the results of block-sparse
decoding for dictionaries with blocks of varying sizes as well. The
generated dictionaries contain $15$ blocks of size $4$ and $20$
blocks of size $3$, with $N=60$ and $K=2N=120$. In this example, we
set $k=2$ and $M=14$. The results are shown as a function of
$\alpha$ in Fig.~\ref{fig:WCM_s} for dictionaries with normally
distributed entries and in Fig.~\ref{fig:WCM_s_DCT} for structured
dictionaries.

\section{Conclusions}
\label{sec:conclusions}

In this paper, we proposed a framework for the design of a sensing
matrix, assuming that a block-sparsifying dictionary is provided. We
minimize a weighted sum of the total inter-block coherence and the
total sub-block coherence, while attempting to keep the atoms in the
equivalent dictionary as normalized as possible (see \eqref{opt1}).
This objective can be seen as an intuitive extension of \eqref{sap}
to the case of blocks.

While it might be possible to derive a closed form solution to
\eqref{opt1}, we have presented the Weighted Coherence Minimization
algorithm, an elegant iterative solution which is based on the
bound-optimization method. In this method, the original objective is
replaced with an easier to solve surrogate objective in each step.
This algorithm eventually converges to a local solution of
\eqref{opt1}.

Simulations have shown that the best results are obtained when
minimizing mostly the total sub-block coherence. This leads to
equivalent dictionaries with nearly orthonormal blocks, at the price
of a slightly increased total inter-block coherence. The obtained
sensing matrix outperforms the one obtained when using the DS
algorithm \cite{sapiro} to solve \eqref{sap}. This improvement
manifests itself in lower signal reconstruction errors and higher
rates of successful signal classification. When giving equal weight
to the total inter-block coherence and to the total sub-block
coherence, the results are identical to solving \eqref{sap}.
Moreover, both objectives coincide for this specific choice of
$\alpha$, which ignores the existence of a block structure in the
sparse representations of the signal data.

\appendices

\section{Proof of convergence}

The surrogate function $g(G,G^{(n)})$ has been chosen in such a way
as to bound the original objective $f(G)$ from above for every $G$,
and to coincide at $G=G^{(n)}$. Minimizing $g(G,G^{(n)})$ will then
necessarily decrease the value of $f(G)$:
\begin{align*}
&\min_G{g(G,G^{(n)})} \leq {g(G^{(n)},G^{(n)})} = f(G^{(n)}),\\
&f(G^{(n+1)}) \leq {g(G^{(n+1)},G^{(n)})} = \min_G{g(G,G^{(n)})}.
\end{align*}
Formally, according to \cite{BO}, the sequence of solutions
generated by iteratively solving
\begin{equation}\label{BO}
    G^{(n+1)} = \arg\min_G{g(G,G^{(n)})}
\end{equation}
is guaranteed to converge to a local minimum of the original
objective $f(G)$ when the surrogate objective $g(G,G^{(n)})$
satisfies the following three constraints:
\begin{enumerate}
  \item \textbf{Equality at $G=G^{(n)}$:}
\[g(G^{(n)},G^{(n)}) = f(G^{(n)}).\]
  \item \textbf{Upper-bounding the original function:}
\[g(G,G^{(n)})\geq f(G), \forall G.\]
  \item \textbf{Equal gradient at $G=G^{(n)}$:}
\[\nabla g(G,G^{(n)})|_{G=G^{(n)}} = \nabla f(G)|_{G=G^{(n)}}.\]
\end{enumerate}
We next prove that the three conditions hold.

\begin{proof} \textbf{Equality at $G=G^{(n)}$:} This follows from the definition
of $g(G,G^{(n)})$.\newline \indent\textbf{Upper-bounding the
original function:} Let us rewrite both functions
    $g(G,G^{(n)})$ and $f(G)$ using the definition of the Frobenius norm:
\begin{align*}
&g(G,G^{(n)})=\\
&\quad\sum_{i,j}\sum_{m,n}\left[\frac{1}{2}((G-h_\eta(G^{(n)}))[i,j]^m_n)^2\right.\\
&\quad\quad\quad+(1-\alpha)((G-h_\mu(G^{(n)}))[i,j]^m_n)^2\\
&\left.\quad\quad\quad+\alpha((G-h_\nu(G^{(n)}))[i,j]^m_n)^2\right],
\end{align*}
and
\begin{align*}
&f(G)=\\
&\quad\sum_{i,j}\sum_{m,n}\left[\frac{1}{2}(u_\eta(G)[i,j]^m_n)^2+(1-\alpha)(u_\mu(G)[i,j]^m_n)^2\right.\\
&\left.\quad\quad\quad+\alpha(u_\nu(G)[i,j]^m_n)^2\right].
\end{align*}
The following observations prove that each of the terms in
$g(G,G^{(n)})$ is larger than or equal to its counterpart in $f(G)$,
and therefore $g(G,G^{(n)})\geq f(G)$:
\begin{align*}
    u_\eta(G)[i,j]^m_n &= \left\{
                         \begin{array}{ll}
                           G[i,j]^m_n-1, &\hbox{$i=j, m = n$;} \\
                           0, &\hbox{else.}
                         \end{array}\right.\\
    (G-h_\eta(G^{(n)}))[i,j]^m_n &= \left\{
                         \begin{array}{ll}
                           G[i,j]^m_n-1, \hbox{ \ $i=j, m = n$;} \\
                           (G-G^{(n)})[i,j]^m_n, \hbox{ \ else.}
                         \end{array}\right.\\
    u_\mu(G)[i,j]^m_n &= \left\{
                         \begin{array}{ll}
                           G[i,j]^m_n, &\hbox{$i\neq j$;} \\
                           0,  &\hbox{else.}
                         \end{array}\right.\\
    (G-h_\mu(G^{(n)}))[i,j]^m_n &= \left\{
                         \begin{array}{ll}
                           G[i,j]^m_n, &\hbox{$i\neq j$;} \\
                           (G-G^{(n)})[i,j]^m_n, &\hbox{else.}
                         \end{array}\right.\\
    u_\nu(G)[i,j]^m_n &= \left\{
                         \begin{array}{lll}
                           G[i,j]^m_n, &\hbox{$i=j, m \neq n$;} \\
                           0,  &\hbox{else.}
                         \end{array}\right.\\
    (G-h_\nu(G^{(n)}))[i,j]^m_n &= \left\{
                         \begin{array}{lll}
                           G[i,j]^m_n,  \hbox{ \ $i=j, m \neq n$;} \\
                           (G-G^{(n)})[i,j]^m_n, \hbox{ \ else.}
                         \end{array}\right.
\end{align*}
\indent\textbf{Equal gradient at $G=G^{(n)}$:} We calculate the
gradient of $g(G,G^{(n)})$ and $f(G)$:
\begin{align*}
&\nabla g(G,G^{(n)})=\\
&\quad2\left[\frac{1}{2}(G-h_\eta(G^{(n)}))+(1-\alpha)(G-h_\mu(G^{(n)}))\right.\\
&\left.\quad+\alpha(G-h_\nu(G^{(n)}))\right],\\
&\nabla f(G)=2\left[\frac{1}{2}u_\eta(G)+(1-\alpha)u_\mu(G)+\alpha
u_\nu(G)\right].
\end{align*}
When substituting $G=G^{(n)}$ we obtain:
\begin{align*}
&\nabla g(G,G^{(n)})|_{G=G^{(n)}}=\nabla f(G)|_{G=G^{(n)}}\\
&=2(\frac{1}{2}u_\eta(G^{(n)})+(1-\alpha)u_\mu(G^{(n)})+\alpha
u_\nu(G^{(n)})).
\end{align*}
Therefore, the gradients of both objectives coincide at $G=G^{(n)}$.
This completes the convergence proof.
\end{proof}

\section{Proof of Proposition 1}

\begin{proof}
In order to minimize $g(G,G^{(n)})$, we rewrite the problem in an
alternative form:
\begin{align}
 &\min_A g(G,\cdot)=\nonumber\\
&\min_A\textrm{tr}\left(\frac{3}{2}G'G-2G'\left[\frac{1}{2}h_\eta(\cdot)+ (1-\alpha)h_\mu(\cdot)+\alpha h_\nu(\cdot)\right]\right)\nonumber\\
=&\min_A\textrm{tr}(E'E E'E -2E'E h_t(\cdot))\nonumber\\
=&\min_A\textrm{tr}(E E'E E'-2E h_t(\cdot)E')\nonumber\\
=&\min_A\textrm{tr}(ADD'A'ADD'A'-2AD h_t(\cdot)D'A'),
 \label{GMM2}
\end{align}
where $h_t(\cdot)$ is defined in \eqref{h_t}. Let $U\Lambda U'$ be
the eigenvalue decomposition of $DD'$ and define $\Gamma_{M\times
N}= AU\Lambda^{1/2}$. Substituting into \eqref{GMM2} yields:
\begin{align}
&\min_A g(G,\cdot)=\nonumber\\
&\min_A\textrm{tr}(\Gamma\Gamma'\Gamma\Gamma'-2\Gamma\Lambda^{-1/2}U'Dh_t(\cdot)D'U\Lambda^{-1/2}\Gamma')\nonumber\\
=&\min_A\|\Gamma'\Gamma-\tilde{h}_t(\cdot)\|_F^2, \label{GMM3}
\end{align}
where $\tilde{h}_t(\cdot)\equiv
\Lambda^{-1/2}U'Dh_t(\cdot)D'U\Lambda^{-1/2}$. According to
\eqref{GMM3}, the surrogate objective $g(G,G^{(n)})$ can be
minimized in closed form by finding the top $M$ components of
$\tilde{h}_t(G^{(n)})$. Let $\Delta_M$ be the top $M$ eigenvalues of
$\tilde{h}_t(G^{(n)})$ and $V_M$ the corresponding $M$ eigenvectors.
Then, \eqref{GMM3} is solved by setting $\Gamma=\Delta_M^{1/2}V_M'$.
Note that this solution is not unique, since $\Gamma$ can be
multiplied on the left by any unitary matrix. Finally, the optimal
sensing matrix is given by $A^{(n+1)} = \Gamma \Lambda^{-1/2}
U'=\Delta_M^{1/2}V_M'\Lambda^{-1/2}U'$. The resulting Gram matrix
$G^{(n+1)}$ is not influenced by the multiplication of $A^{(n+1)}$
on the left by a unitary matrix. Therefore, the WCM algorithm is not
affected by the choice of $A^{(n+1)}$.
\end{proof}

\section*{Acknowledgements}

The research of Lihi Zelnik-Manor is supported by Marie Curie
IRG-208529.

\newpage
\bibliographystyle{IEEEbib}
\bibliography{refs}

\end{document}